\documentclass[11pt,a4paper]{article}
\usepackage{jheppub}

\usepackage{slashed,color,amsmath,amssymb,mathrsfs}
\usepackage{graphicx}
\usepackage{longtable}
\usepackage{array}
\usepackage{autobreak}
\usepackage{booktabs}
\usepackage{cancel}
\usepackage{bm}
\usepackage{physics}
\usepackage{accents}

\allowdisplaybreaks

\newcommand{\la}{\langle}
\newcommand{\ra}{\rangle}

\newcommand{\fp}{f_+}

\newcommand{\fm}{f_-}

\newcommand{\chip}{\chi_+}

\newcommand{\chim}{\chi_-}

\newcommand{\ahe}{{\hat{\alpha}_\perp}}

\newcommand{\ahp}{{\hat{\alpha}_\parallel}}

\newcommand{\As}{\hat{\mathscr{A}}}

\newcommand{\Hs}{\hat{\mathscr{H}}}

\newcommand{\Vs}{\hat{\mathscr{V}}}

\newcommand{\Vt}{V}

\newcommand{\chiph}{\hat{\chi}_+}

\newcommand{\chimh}{\hat{\chi}_-}

\begin{document}


\title{High order chiral Lagrangians with vector mesons in different approaches}

\author[a,b]{Wei Guo}
\author[a,c]{Qin-He Yang}
\author[a]{Shao-Zhou Jiang}

\affiliation[a]{Guangxi Key Laboratory for Relativistic Astrophysics, School of Physical Science and Technology, Guangxi University,  Nanning 530004, China}
\affiliation[b]{Department of Physics, Beijing Normal University,
Beijing 100875, China},
\affiliation[c]{School of Physics and Electronics, Hunan University, Changsha 410082, China}

\emailAdd{guow@mail.bnu.edu.cn}
\emailAdd{yqh@st.gxu.edu.cn}
\emailAdd{jsz@gxu.edu.cn}

\abstract{
The chiral Lagrangians with vector mesons are constructed in different approaches, including the next-to-leading order Lagrangian in the vector-field approach, the next-to-next-to-leading order Lagrangians in the tensor-field and the hidden local symmetry approaches. Some redundant terms are found at the next-to-leading order in the tensor-field and the hidden local symmetry approaches. The corresponding relations between the next-to-next-to-leading order pseudoscalar mesonic low-energy constants and the ones in the hidden local symmetry approach are obtained at tree level. The equivalence between the tensor-field approach and the hidden local symmetry approach is also discussed.

}

\keywords{Effective Field Theory, Chiral Lagrangians, Vector mesons}

\maketitle

\date{\today}

\flushbottom

\section{
 Introduction}

It is well known that quantum chromodynamics (QCD) is the fundamental theory to describe the strong interactions. However, at present, QCD usually could not give a very precise prediction in the low-energy region analytically because of the non-perturbative effects. Owing to the chiral symmetry and its spontaneous breaking, one can study the low-energy QCD involving low-energy pseudoscalar mesons with an effective theory at the hadron level. It avoids considering the complicated interactions between quarks and gluons. This approach is called chiral perturbation theory (ChPT) \cite{callan_structure_1969,coleman_structure_1969,weinberg_phenomenological_1979,gasser_chiral_1984,gasser_chiral_1985}. ChPT is an effective field theory of QCD in the low energy, while QCD is an ultraviolet completion of ChPT in the high energy. In ChPT, the chiral Lagrangian includes all effective interactions, which are invariant under the Lorentz transformation, chiral rotation, charge conjugation transformation and parity transformation. Instead of the strong coupling constant, ChPT is expanded by the typical scale of momentum ($p$). According to Weinberg’s power-counting rule, the tree and the loop diagrams are counted as some suitable powers of $p/\Lambda$, where $\Lambda$ is related to the chiral symmetry breaking scale. Hence, the calculation of a physical quantity at a certain degree of accuracy is according to a finite number of interactions. At present, the pseudoscalar mesonic chiral Lagrangians have been constructed up to the $\mathcal{O}(p^8)$ order in both $SU(2)$ and $SU(3)$ situations \cite{gasser_chiral_1984,gasser_chiral_1985,fearing_extension_1996,bijnens_mesonic_1999,bijnens_anomalous_2002,bijnens_order_2018}.

ChPT can be extended to the higher energy range and incorporate other particles, such as vector mesons. Vector mesons are very special and they have their own features. The masses of vector mesons are between pseudoscalar mesons and baryons, which contribute to an inconsistent power counting rule. Some literature treats them as light degrees of freedom \cite{prades_massive_1994,Rosell:2004mn,Rosell:2006dt,lutz_radiative_2008,leupold_hadronic_2009,Rosell:2009yb,kampf_renormalization_2010,Pich:2010sm,danilkin_causality_2011,terschlusen_electromagnetic_2012,danilkin_photon-fusion_2013,gao_parton_2014,guo_light_2019}, while other literature treats them as heavy degrees of freedom \cite{jenkins_chiral_1995,bijnens_electromagnetic_1996,bijnens_chiral_1998,djukanovic_path_2010,bruns_chiral_2013}. There exists more than one approach to represent them, such as the traditional vector-field (also called matter field or Proca field) approach \cite{weinberg_nonlinear_1968,ecker_chiral_1989}, the antisymmetric tensor-field approach \cite{gasser_chiral_1984,ecker_chiral_1989,ecker_role_1989}, the hidden local symmetry (HLS) approach \cite{bando_is_1985,bando_vector_1985,bando_composite_1985,fujiwara_non-abelian_1985,tanabashi_chiral_1993,tanabashi_formulations_1996,harada_wilsonian_2001,harada_hidden_2003,ma_hidden_2005} and the massive Yang-Mills approach \cite{gasiorowicz_effective_1969,kaymakcalan_non-abelian_1984,gomm_anomalous_1984,meissner_low-energy_1988}. In this paper, we will discuss the first three approaches, because the massive Yang-Mills approach is rarely used at present. These approaches are not independent and they have equivalent relations with each other. The vector-field approach is the most traditional one, whose relations with the others have been given in Refs. \cite{ecker_chiral_1989,bijnens_tensor_1996,birse_effective_1996,harada_hidden_2003,kampf_different_2007}. In the tensor-field approach, a vector field is represented by an antisymmetric tensor. All tensor fields transform homogeneously under the chiral symmetry \cite{ecker_chiral_1989,ecker_role_1989,bijnens_tensor_1996,kampf_different_2007}. The vector- and the tensor-field approaches have so much in common. Some literature discusses them simultaneously \cite{ecker_chiral_1989,bijnens_tensor_1996, kampf_different_2007}. The tensor-field approach has its advantages \cite{Jenkins:1972nj,chizhov_theory_2011}. First, the antisymmetric tensor field is the direct generalization to spin one of the Dirac equation for spin half particles. Second, the tensor-field approach corresponds to the $(1,0)\oplus (0,1)$ representation of the Lorentz group. It contains six degrees of freedom, which can be separated into two spin-one degrees of freedom. Nevertheless, the vector-field approach corresponds to the $(\frac{1}{2},\frac{1}{2})$ representation. It needs a subsidiary condition to remove the unwanted spin-zero component. Third, the tensor-field Lagrangian has a simpler form than that in the vector field. The relations between these two approaches can be found in Refs. \cite{ecker_chiral_1989,birse_effective_1996,bijnens_tensor_1996,tanabashi_formulations_1996,kampf_different_2007}. For the HLS approach, an artificial local symmetry is introduced to the nonlinear sigma model by choosing special field variables. This symmetry can be removed by choosing some specific gauge conditions, e.g. unitary gauge. In fact, the gauge symmetry is not a symmetry but a simply redundancy of the description. HLS approach also has relations with the other approaches \cite{ecker_chiral_1989,pallante_anomalous_1993,tanabashi_formulations_1996,harada_hidden_2003}. The massive Yang-Mills approach is a special gauge of the generalized HLS with a particular parameter choice. The relations between the massive Yang-Mills approach and the other approaches have been given in Refs. \cite{Meissner:1986tc,Yamawaki:1986zz,Golterman:1986cz,ecker_chiral_1989}.

For these approaches, there exist two different equivalent relations \cite{pallante_anomalous_1993,kampf_different_2007}. One means that different approaches have the same structure without integrating vector mesons \cite{tanabashi_formulations_1996,harada_hidden_2003}, while the other one means that different approaches give equivalent relations after integrating the vector fields \cite{ecker_chiral_1989,bijnens_tensor_1996,kampf_different_2007}. The former one is at the level of the resonance Lagrangians and the latter one is at the level of the pseudoscalar mesonic chiral Lagrangians. The first one implies the second one, but not vice versa. After integrating the vector fields and comparing with the pseudoscalar mesonic chiral Lagrangians \cite{gasser_chiral_1984,gasser_chiral_1985,bijnens_chiral_2007,bijnens_mesonic_1999}, the equivalences among these approaches at the level of the pseudoscalar mesonic chiral Lagrangian can be accounted for \cite{ecker_chiral_1989,bijnens_tensor_1996,kampf_different_2007}. A new method called the first order formalism gives resonance contributions to the low-energy constants (LECs) in ChPT at the next-to-next-to-leading order (NNLO) by integrating vector fields \cite{kampf_different_2007}. This approach accounts for the equivalence between the vector-field and the tensor-field approach at the NNLO.

At present, although there exist a lot of researches based on the ChPT with vector mesons, most of them regard vector mesons as resonance states to study the pseudoscalar mesons \cite{RuizFemenia:2003hm,Cirigliano:2004ue,Cirigliano:2005xn,cirigliano_towards_2006,Dai:2019lmj}. In this case, the low-order chiral Lagrangian is enough. On the other hand, some researches study the vector mesons themselves, such as the vector meson masses \cite{bijnens_electromagnetic_1996,Bijnens:1997ni,bruns_infrared_2005,Djukanovic:2009zn,Bruns:2013tja,Kawaguchi:2015gpt,Bavontaweepanya:2018yds}, their electromagnetic properties \cite{danilkin_photon-fusion_2013,Unal:2019eum}, vector meson decay constants \cite{Bijnens:1998di}, vector meson radiative decays \cite{lutz_radiative_2008}, vector meson scattering with the other high energy particles \cite{ma_hidden_2005,Zhou:2014ila,terschlusen_electromagnetic_2012,Djukanovic:2018pep,Djukanovic:2004mm} and so on. Some of them study vector mesons at the next-to-leading order (NLO). The NNLO research is also mentioned. At present, the vector mesonic chiral Lagrangians have been obtained to the NLO in the tensor-field and the HLS approaches \cite{bando_is_1985,ecker_chiral_1989,tanabashi_chiral_1993,cirigliano_towards_2006}. Generally, if the study is on the vector mesons themselves, the low-order chiral Lagrangian may not give the results in enough precision. In this paper, we will construct the chiral Lagrangians in these three approaches including the normal and the anomalous parts to a given order. For the vector-field approach, the Lagrangian is constructed to the NLO. The Lagrangian is constructed to the NNLO in the tensor-field and the HLS approaches. A short discussion about the strong equivalence between the tensor-field and the HLS approaches is also given. Although there already exist researches on $n$-vector-meson interactions, most researches only consider one-vector-meson vertices. The $n$-vector-meson interactions in the vector-field and the tensor-field approaches are more complicated. Hence, only one-vector-meson vertex is mentioned in the vector-field and the tensor-fields approaches. The high-order chiral Lagrangians contain a lot of terms, but this is not a serious problem. Only a few terms have an impact on a special problem. With the rapid development of technologies, more and more software arises. Some computations can be done by using the computer algebra system.

This paper only gives a discussion at tree level. If one wants to discuss the loop diagrams, the chiral Lagrangians themselves are not enough, because the vector mesons upset the power counting \cite{bruns_infrared_2005}. The loop diagrams lead to large imaginary parts. These parts violate the power counting and can not be cancelled by the counterterms or absorbed in the renormalized parameters as the usual renormalization scheme. The only method to solve this problem at present is the complex-mass scheme \cite{Denner:1999gp,Denner:2005fg,Denner:2006ic,Actis:2006rc,Actis:2008uh,Djukanovic:2009zn,Djukanovic:2009gt,Denner:2014zga,Djukanovic:2015gna,Gegelia:2016xcw,Gegelia:2016pjm}. In this scheme, the masses and the widths of vector mesons are counted as $\mathcal{O}(p^0)$ and $\mathcal{O}(p^1)$, respectively. For a propagator of vector mesons, it is counted as $\mathcal{O}(p^0)$ if it carries large external momenta, otherwise, it is counted as $\mathcal{O}(p^1)$. The derivative acting on vector mesons is counted as $\mathcal{O}(p^0)$. The power counting related to the pseudoscalar mesons is the same as that in the pseudoscalar mesonic ChPT. The power-counting-violating terms can be cancelled by the counterterms, because the parameters in the counterterms also become complex. Hence, the power counting still holds in the complex-mass scheme and the chiral expansion also works at loop level.

This paper is organized as follows. Section \ref{sec:_chiral_su_n_} gives a short review on the pseudoscalar and vector mesons in ChPT. Sections \ref{vfa} to \ref{ha} give the methods of constructing the chiral Lagrangians with vector mesons in the vector-field, the tensor-field and the HLS approaches, respectively. Section \ref{rcl} gives the results of chiral Lagrangians. Section \ref{rh} gives the relations of the LECs between the HLS Lagrangian and the pseudoscaler mesonic chiral Lagrangian at tree level. In Sec. \ref{sec:equivalence_of_different_approachs}, we give a discussion about the equivalence between the HLS and the tensor-field approaches. Section \ref{sec:conclusions_and_discussions} is a short summary.

\section{Mesons in chiral perturbation theory} \label{sec:_chiral_su_n_}
In QCD, the Lagrangian with external sources is
\begin{align}
\mathscr{L}_{\mathrm{QCD}}=\mathscr{L}_{\mathrm{QCD}}^{0}+\bar{q}(\slashed{v}+\slashed{a}\gamma_5-s+ip\gamma_5)q ,
\end{align}
where $\mathscr{L}_{\mathrm{OCD}}^{0}$ is the original QCD Lagrangian. $q$ denotes the quark fields. $v^{\mu}$, $a^{\mu}$, $s$ and $p$ represent vector, axial-vector, scalar and pseudoscalar sources, respectively. For convenience, the tensor source and the $\theta$ term are ignored. To compare with the pseudoscalar mesonic chiral Lagrangian, $a^{\mu}$ is considered traceless, and $v^{\mu}$ is only traceable in the two-flavor anomalous part, it is traceless in the other cases.

If the masses of the light quarks are ignored, the QCD Lagrangian exists a global $SU(N_f)_L\times SU(N_f)_R$ chiral symmetry, where $N_f=$ 2 or 3. Chiral symmetry spontaneously breaks into $SU(N_f)_V$ symmetry. The related pseudoscalar Goldstone bosons are regarded as the lowest pseudoscalar mesons $U$. $U$ transforms as $U\rightarrow g_L U g_R^{\dagger}$ under the chiral rotation, where $g_R$ and $g_L$ denote the group elements in $SU(N_f)_L$ and $SU(N_f)_R$ rotations, respectively. For convenience, another field $u$ is introduced to describe the meson fields, where $u^2=U$. $u$ transforms as $u\rightarrow g_L u h^{\dagger}=hug_R^\dag$ under the chiral rotation, where $h$ is a compensator field which is dependent on the meson fields. In this form, it is more convenient to construct the chiral Lagrangians and we will adopt it.

The vector-meson octet is represented by a $3\times 3$ matrix in the vector-field approach
\begin{align}
V^{\mu}=\left( \begin{array}{ccc}{\displaystyle\frac{\rho^{0}}{\sqrt{2}}+\frac{\omega_8}{\sqrt{6}}  } & {\rho^{+}} & {K^{*+}} \\
{\rho^{-}} & \displaystyle{-\frac{\rho^{0}}{\sqrt{2}}+\frac{\omega_8}{\sqrt{6}} } & {K^{* 0}} \\
{K^{*-}} & {\overline{K}^{* 0}} & \displaystyle{-\frac{2 \omega_8}{\sqrt{6}}}
\end{array}\right)^{\mu}.\label{dV}
\end{align}
It transforms to $h V^{\mu}h^{\dagger}$ under the chiral rotation. In the tensor-field approach, the vector field $V^\mu$ is replaced by an antisymmetric tensor $W^{\mu\nu}$ and all elements in matrix $V^{\mu}$ are replaced by the corresponding antisymmetric tensor fields. The HLS approach is a little more complicated and it will be discussed in Sec. \ref{ha}.

We mainly consider the vector-meson octet in this work. Most discussions focus on the octet unless specified. If one wants to consider the fine effects of $\rho^0-\omega$ mixing \cite{Urech:1995ry} or $\omega-\phi$ mixing \cite{Kucukarslan:2006wk}, the vector-meson singlet $\omega_0$ needs to be added to Eq. \eqref{dV}. The ideal mixing causes
\begin{align}
\begin{pmatrix}
\omega^\mu\\
\phi^\mu
\end{pmatrix}
=\begin{pmatrix}
\sqrt{\frac{1}{3}} & \sqrt{\frac{2}{3}}\\
-\sqrt{\frac{2}{3}} & \sqrt{\frac{1}{3}}
\end{pmatrix}
\begin{pmatrix}
\omega_8^\mu\\
\omega_0^\mu
\end{pmatrix}.
\end{align}
The more precise $\rho^0-\omega$ ($\omega-\phi$) mixing need to consider $\rho^0-\omega$ ($\phi-\omega$) interactions. This work only considers one-vector-meson vertices and does not contain these interactions. There are two methods to introduce the vector singlet. One method is to extend the $SU(3)$ presentation to $U(3)$, i.e.
\begin{align}
V^\mu\to V^\mu+\frac{\omega_0^\mu}{\sqrt{3}} I_3,\label{V9}
\end{align}
where $I_3$ is the $3\times 3$ unit matrix. The other method does not assume the $U(3)$ symmetry and considers $\phi^\mu$ ($\omega^\mu$ in two-flavor case) as a building block in the calculation.

In this paper, we do not assume the isospin symmetry. If one wants to consider the isospin breaking, such as the mass difference between $\rho^0$ and $\rho^\pm$, the mixing among $\rho^0$, $\omega$ and $\phi$ needs to be considered and $s$ should be substituted into $\mathrm{diag}(m_u,m_d,m_s)$ in the calculation \cite{Bijnens:1996nq,Bijnens:1997ni}.

The following three sections will discuss how to construct the chiral Lagrangians with vector mesons in the different approaches, respectively. In this paper, we only consider one-vector-meson vertices in the vector-field and the tensor-field approaches.

\section{Vector-field approach}\label{vfa}
\subsection{Building blocks and transformation properties}\label{bbp}
Besides the vector meson fields, the building blocks in the vector-field approach are the same as those in the pseudoscalar mesonic chiral Lagrangians \cite{bijnens_mesonic_1999,bijnens_chiral_2007},
\begin{align}
u^{ \mu } & = i \left\{ u ^ { \dagger } \left( \partial^{ \mu } - i r^{ \mu } \right) u - u \left( \partial^{ \mu } - i \ell^{ \mu } \right) u ^ { \dagger } \right\}, \notag\\
\chi _ { \pm } & = u ^ { \dagger } \chi u ^ { \dagger } \pm u \chi ^ { \dagger } u, \notag\\
f_{+}^{\mu \nu} &=u F_{L}^{\mu \nu} u^{\dagger} + u^{\dagger} F_{R}^{\mu \nu} u,\notag\\
f_{-}^{\mu \nu} &=u F_{L}^{\mu \nu} u^{\dagger} - u^{\dagger} F_{R}^{\mu \nu} u=-\nabla^{\mu} u^{\nu}+\nabla^{\nu} u^{\mu} ,\notag\\
h^{\mu \nu} &=\nabla^{\mu} u^{\nu}+\nabla^{\nu} u^{\mu},\label{bb}
\end{align}
where
$r^{\mu}=v^{\mu}+a^{\mu}$, $\ell^{\mu}=v^{\mu}-a^{\mu}$, $\chi=2 B_0(s+i p)$, $F_{R}^{\mu \nu}=\partial^{\mu} r^{\nu}-\partial^{\nu} r^{\mu}-i\left[r^{\mu}, r^{\nu}\right],F_{L}^{\mu \nu}=\partial^{\mu} \ell^{\nu}-\partial^{\nu} \ell^{\mu}-i\left[\ell^{\mu}, \ell^{\nu}\right]$ and $B_0$ is a constant related to the quark condensate. These building blocks have the same chiral rotation properties as $V^{\mu}$, i.e. $X \rightarrow h X h^\dagger$, where $X$ is any building block. Their covariant derivatives are
\begin{align}
\nabla^{\mu} X=\partial^{\mu} X+\left[\Gamma^{\mu}, X\right],
\end{align}
where $\Gamma_{\mu}$ is the chiral connection
\begin{align}
\Gamma^{\mu}=\frac{1}{2}\left\{u^{\dagger}\left(\partial^{\mu}-i r^{\mu}\right) u+u\left(\partial^{\mu}-i \ell^{\mu}\right) u^{\dagger}\right\}.
\end{align}

Generally, in order to construct the vector mesonic chiral Lagrangians, $V^\mu$, $X$ and their derivatives should be considered. Some literature also introduces a building block $\hat{V}^{\mu\nu}=\nabla^{\mu}V_{\nu}-\nabla^{\nu}V^{\mu}$ \cite{ecker_chiral_1989,kampf_different_2007}. However, this building block is not necessary. The covariant derivatives in front of $V^{\mu}$ can be moved to the other building blocks by partial integration. This will be discussed in Eq. \eqref{pir}. Hence, besides the kinetic term, we do not consider this building block.

In addition to chiral symmetry, the chiral Lagrangians also need to be invariant under the transformations of parity ($P$), charge conjugation ($C$), and Hermitian conjugation (h.c.). The transformation properties of these building blocks have been given in a lot of literature \cite{bijnens_mesonic_1999,bijnens_chiral_2007,ecker_chiral_1989,ecker_role_1989}. For convenience, these properties are presented in Table \ref{blbt}.
\begin{table*}[!h]
\caption{\label{blbt}Chiral dimension (Dim), parity ($P$), charge conjugation ($C$) and hermiticity (h.c.) of the building blocks in the vector-field and the tensor-field approaches.}
\begin{center}
\begin{tabular}{ccccc}
\hline\hline
& Dim &                $P$                 &                $C$                &               h.c.                \\
\hline
$u^{\mu}$             &  1  &             $-u_{\mu}$             &           $(u^{\mu})^T$           &             $u^{\mu}$             \\
$h^{\mu\nu}$            &  2  &           $-h_{\mu\nu}$            &         $(h^{\mu\nu})^T$          &           $h^{\mu\nu}$            \\
$\chi_{\pm}$            &  2  &          $\pm\chi_{\pm}$           &         $(\chi_{\pm})^T$          &         $\pm \chi_{\pm}$          \\
$f_{\pm}^{\mu\nu}$         &  2  &        $\pm f_{\pm\mu\nu}$         &    $\mp (f_{\pm}^{\mu\nu})^T$     &        $ f_{\pm}^{\mu\nu}$        \\
$V^{\mu}$             &  0  &                $V_{\mu}$                &           $-(V^{\mu})^T$            &             $V^{\mu}$              \\
$W^{\mu\nu}$             &  0  &                $W_{\mu\nu}$                &           $-(W^{\mu\nu})^T$            &             $W^{\mu\nu}$              \\
$\varepsilon^{\mu\nu\lambda\rho}$ &  0  & $-\varepsilon_{\mu\nu\lambda\rho}$ & $\varepsilon^{\mu\nu\lambda\rho}$ & $\varepsilon^{\mu\nu\lambda\rho}$ \\
\hline\hline
\end{tabular}
\end{center}
\end{table*}

The vector meson masses are
\begin{align}
\rho^\pm, \rho^0(775\mathrm{MeV}),\quad \omega(782\mathrm{MeV}),\quad K^{*\pm}(892\mathrm{MeV}),\quad K^{*0}, \bar{K}^{*0}(896\mathrm{MeV}).
\end{align}
Their masses are between the pseudoscalar mesons and the baryons masses. They can be regarded as either light or heavy degrees of freedom. In this paper, we count $V^\mu=\mathcal{O}(p^0)$. The chiral dimensions of the other building blocks are the same as those in the pseudoscalar mesonic chiral Lagrangians \cite{leupold_towards_2012,harada_hidden_2003}. Table \ref{blbt} gives a summary about these power counting rules.
In addition, each covariant derivative on these building blocks contributes an additional $\mathcal{O}(p^1)$ order. For the anomalous part, Levi-Civita tensor $\varepsilon^{\mu\nu\lambda\rho}$ is also needed. The above power counting works fine at tree level. If one wants to consider the loop contributions, the complex-mass scheme should be introduced as the discussion in the introduction.

\subsection{Constraint relations}\label{vcr}
Chiral Lagrangian needs to satisfy all the symmetries discussed above. It is easy to construct a complete set which satisfies all of these symmetries, but the terms in this set are generally not linearly independent. In order to get a linearly independent chiral Lagrangian, the following linear constraint relations need to be considered \cite{fearing_extension_1996,bijnens_mesonic_1999,bijnens_anomalous_2002,bijnens_order_2018,ecker_chiral_1989,ecker_role_1989}.

\begin{enumerate}
  \item Partial integration. Because the derivative acting on the whole Lagrangian can be discarded, it leads to
  \begin{align}
  \la\nabla^{\mu} A B \cdots\ra\la C \cdots\ra+\la A \nabla^{\mu} B \cdots\ra\la C \cdots\ra+\la A B \cdots\ra\la\nabla^{\mu} C \cdots\ra+\text{other terms}=0, \label{pir}
  \end{align}
  where $A$, $B$, $C$ are some building blocks, ``$\la\cdots\ra$'' represents the flavor trace and ``$\cdots$'' denotes one or more possible building blocks. From this relation, one of the terms on the left-hand side in Eq. \eqref{pir} is not independent and it can be discarded. For convenience, using this relation, the covariant derivatives in front of $V^\mu$ are moved to the other building blocks. Hence, the terms containing $\hat{V}^{\mu\nu}$ or $\nabla^\mu V^\nu$ will not appear.

  \item Equation of motions (EOMs). Only the leading order EOMs are needed for constructing the chiral Lagrangians \cite{bijnens_mesonic_1999}. For pseudoscalar mesons, the leading order EOM is
  \begin{align}
  \nabla^{\mu} u_{\mu}=\frac{i}{2}\left(\chi_{-}-\frac{1}{N_f}\left\la\chi_{-}\right\ra\right).
  \end{align}
  With this relation, $\nabla^{\mu}u_{\mu}$ can be replaced by other factors and it does not exist in the chiral Lagrangians.

  For vector mesons, the leading order Lagrangian is \cite{ecker_chiral_1989}
  \begin{align}
  \mathcal{L}_{2}=-\frac{1}{4}\left\la \hat{V}_{\mu \nu} \hat{V}^{\mu \nu}-2 M_V^{2} V_{\mu} V^{\mu}\right\ra.
  \end{align}
  This Lagrangian leads to Klein-Gordon equation and a subsidiary condition,
  \begin{align}
  (\nabla_{\nu}\nabla^{\nu}+M_V^2) V^{\mu}\doteq 0,\quad \quad \nabla_\mu V^\mu\doteq 0, \label{eomv}
  \end{align}
  where the symbol ``$\doteq$''  means that both sides are equal if other forms are ignored. The above equations mean $\nabla_{\nu}\nabla^{\nu}V^{\mu}$ and $\nabla_\mu V^\mu$ can be transformed into the other forms. Combining with partial integration, Eq. \eqref{eomv} can be written as
  \begin{align}
  V^{\mu}\nabla^{\rho}\nabla_{\rho}O_{\mu}\doteq 0,\quad  V^{\mu}\nabla_{\mu}O\doteq0,
  \end{align}
  where $O$ $(O_{\mu})$ represents the product of any set of the building blocks with suitable indices. The ignored parts have no impact on the structure of chiral Lagrangians.

  \item  Bianchi identity. With the definitions in Sec. \ref{bbp}, the following Bianchi identity is obtained
  \begin{align}
  \nabla^{\mu} \Gamma^{\nu \rho}+\nabla^{\nu} \Gamma^{\rho \mu}+\nabla^{\rho} \Gamma^{\mu \nu}=0,
  \end{align}
  where for any building block $X$,
  \begin{align}
  \left[\nabla^{\mu}, \nabla^{\nu}\right] X &=\left[\Gamma^{\mu \nu}, X\right], \\
  \Gamma^{\mu \nu} &=\frac{1}{4}\left[u^{\mu}, u^{\nu}\right]-\frac{i}{2} f_{+}^{\mu \nu}.
  \end{align}
  Two special cases are
  \begin{align}
  \nabla^{\mu} f_{-}^{\nu \alpha}+\nabla^{\nu} f_{-}^{\alpha \mu}+\nabla^{\alpha} f_{-}^{\mu \nu}&=\frac{i}{2}\left(\left[f_{+}^{\mu \nu}, u^{\alpha}\right]+\left[f_{+}^{\nu \alpha}, u^{\mu}\right]+\left[f_{+}^{\alpha \mu}, u^{\nu}\right]\right),\\
  \nabla^{\mu} f_{+}^{\nu \alpha}+\nabla^{\nu} f_{+}^{\alpha \mu}+\nabla^{\alpha} f_{+}^{\mu \nu}&=\frac{i}{2}\left(\left[f_{-}^{\mu \nu}, u^{\alpha}\right]+\left[f_{-}^{\nu \alpha}, u^{\mu}\right]+\left[f_{-}^{\alpha \mu}, u^{\nu}\right]\right).
  \end{align}

  \item Schouten identity. For the anomalous parts, Levi-Civita tensor $\epsilon^{\mu \nu \alpha \beta}$ appears. Because there does not exit any total antisymmetric fifth order tenor in the four-dimensional spacetime, Schouten identity indicates that
  \begin{align}
  S^{\gamma} \epsilon^{\mu \nu \alpha \beta}-S^{\mu} \epsilon^{\gamma \nu \alpha \beta}-S^{\nu} \epsilon^{\mu \gamma \alpha \beta}-S^{\alpha} \epsilon^{\mu \nu \gamma \beta}-S^{\beta} \epsilon^{\mu \nu \alpha \gamma}=0,
  \end{align}
  where $S^{\mu}$ is any operator.

  \item Cayley-Hamilton relations. All building blocks are $N_f\times N_f$ matrix in the flavor space. Cayley-Hamilton theorem gives relations among $N_f\times N_f$ matrices.
  For SU(2),
  \begin{align}
  \{A, B\}=A\la B\ra+ B\la A\ra+\la A B\ra-\la A\ra\la B\ra,\label{eq.240}
  \end{align}
  where $A$ and $B$ are $2\times 2$ any matrices.

  For SU(3),
  \begin{align}
  \begin{array}{l}{A B C+A C B+B A C+B C A+C A B+C B A-A B\la C\ra-} \\ {\quad-A C\la B\ra- B A\la C\ra- B C\la A\ra- C A\la B\ra- C B\la A\ra- A\la B C\ra-} \\ {\quad-B\la A C\ra- C\la A B\ra-\la A B C\ra-\la A C B\ra+ A\la B\ra\la C\ra+ B\la A\ra\la C\ra+} \\ {\quad+C\la A\ra\la B\ra+\la A\ra\la B C\ra+\la B\ra\la A C\ra+\la C\ra\la A B\ra-\la A\ra\la B\ra\la C\ra= 0},\end{array}\label{eq.24}
  \end{align}
  where $A$, $B$ and $C$ are $3\times 3$ any matrices.
\end{enumerate}

\section{Tensor-field approach}
The tensor-field approach is similar to the vector-field approach. The building blocks are almost the same as those in Table \ref{blbt}, except for replacing $V^\mu$ by $W^{\mu\nu}$. The transformation properties and power counting rule of $W^{\mu\nu}$ are also presented in Table \ref{blbt}.

In the tensor-field approach, the linear relations are almost the same as those in the vector-field approach. The only different point is the EOM for vector mesons. The tensor-field $W^{\mu\nu}$ is antisymmetric, $W^{\mu\nu}=-W^{\nu\mu} $. With the methods in Refs. \cite{capri_second_1987,ecker_role_1989,bruns_infrared_2005}, Eq. \eqref{eomv} is replaced by the following two equations,
\begin{align}
(\nabla_{\rho}\nabla^{\rho}+M_{V}^2)W^{\mu\nu}\doteq 0,\quad\nabla^{\lambda} W^{\rho \sigma}+\nabla^{\rho} W^{\sigma \lambda}+\nabla^{\sigma}W^{\lambda\rho}\doteq 0. \label{eomw}
\end{align}
There is another subsidiary condition \cite{capri_second_1987,rosell_quantum_2007,ecker_role_1989,bruns_infrared_2005,bijnens_tensor_1996}
\begin{align}
\nabla_{\rho}W^{\rho\mu}\doteq 0.
\end{align}
For historical reasons, we adopt the convention in Eq. \eqref{eomw}, where the degrees of freedom $W^{ij}$ are frozen but $W^{0i}$ are dynamical. Combining with partial integration, Eq. \eqref{eomw} leads to the following two linear relations
\begin{align}
W_{\mu\nu}\nabla_{\rho}(R^{\mu\nu\rho}+R^{\nu\rho\mu}+R^{\rho\mu\nu})\doteq 0,\quad
W_{\mu\nu}\nabla_{\rho}{\nabla^{\rho}{C^{\mu\nu}}}\doteq0\label{a.4},
\end{align}
where $R^{\mu\nu\rho}$ is any operator with three indices and $C^{\mu\nu}$ is any operator with two indices. The $M_{V}^2$ term is ignored in the last relation.

\section{HLS approach}\label{ha}
The HLS approach is very different from the vector-field approach or the tensor-field approach. This section only gives a short review on it. More details can be found in Refs. \cite{bando_is_1985,bando_composite_1985,bando_vector_1985,bando_nonlinear_1988,tanabashi_chiral_1993,tanabashi_formulations_1996,harada_hidden_2003}.

\subsection{Definitions}
In this approach, vector mesons are considered as gauge fields and their masses come from Brout-Englert-Higgs mechanism. HLS approach is based on the $G_{\text{global}}\times H_{\text{local}}$ symmetry, where $G=SU(N_f)_L\times SU(N_f)_R$ is the global symmetry and $H$ is the HLS. $G_{\text{global}}\times H_{\text{local}}$ symmetry can break into a $H$ symmetry, which is the flavor symmetry. $U$ field in ChPT can be divided into two fields $\xi_L$ and $\xi_R$,
\begin{align}
U=\xi_L^\dagger\xi_R.
\end{align}
The two fields $\xi_L$ and $\xi_R$ satisfy the following transformation properties under the $G_{\text{global}}\times H_{\text{local}}$ symmetry
\begin{align}
\xi_{L,R}(x)\rightarrow h(x)\cdot \xi_{L,R}(x)\cdot g_{L,R}^\dagger,\label{eq.34}
\end{align}
where $h(x)\in H_{\text{local}}$ and $g_{L,R}\in SU(N_f)_L\times SU(N_f)_R$. Two Maurer–Cartan 1-forms are
\begin{align}
\alpha_\perp^{\mu}&=\frac{1}{2i}(\partial^{\mu}\xi_R\cdot\xi_R^{\dagger}-\partial^{\mu}\xi_L\cdot\xi_L^\dagger),\\
\alpha_\parallel^{\mu}&=\frac{1}{2i}(\partial^{\mu}\xi_R\cdot\xi_R^{\dagger}+\partial^{\mu}\xi_L\cdot\xi_L^{\dagger}).
\end{align}
Their chiral transformation properties under the full symmetry are
\begin{align}
\alpha_\perp^{\mu}&\rightarrow h(x)\cdot\alpha_\perp^{\mu}\cdot h^{\dagger}(x),\\
\alpha_\parallel^{\mu}&\rightarrow h(x)\cdot\alpha_\perp^{\mu}\cdot h^{\dagger}(x)-i\partial^{\mu}h(x)\cdot h^{\dagger}(x).
\end{align}
The relations between $\alpha_\perp^{\mu},\alpha_\parallel^{\mu}$ and the building blocks in the vector-(tensor-) field approach are $\alpha_\perp^{\mu}=\frac{1}{2}u^{\mu}$ and $\alpha_\parallel^\mu=i\Gamma^{\mu}$. The covariant derivatives of $\xi_{L,R}$ are
\begin{align}
D^{\mu}\xi_{L}&=\partial^{\mu}\xi_{L}-iK^{\mu}\xi_{L}+i\xi_{L}l^{\mu},\\
D^{\mu}\xi_{R}&=\partial^{\mu}\xi_{R}-iK^{\mu}\xi_{R}+i\xi_{R}r^{\mu},
\end{align}
where $K^{\mu}$ are the corresponding gauge fields in $H_{\text{local}}$. Its chiral transformation property is
\begin{align}
K^{\mu}\rightarrow h(x)\cdot K^{\mu}\cdot h^{\dagger}(x)-i\partial^{\mu}h(x)\cdot h^{\dagger}(x).
\end{align}
With $K^{\mu}$ fields, the covariant 1-forms are
\begin{align}
\ahe^{\mu}=&{\alpha_\perp}^{\mu}=\frac{1}{2i}(D^{\mu}\xi_R\cdot\xi_R^{\dagger}-D^{\mu}\xi_L\cdot\xi_L^{\dagger}),\label{hlsb1}\\
\ahp^{\mu}=&a_\parallel^\mu-K^\mu=\frac{1}{2i}(D^{\mu}\xi_R\cdot\xi_R^{\dagger}+D^{\mu}\xi_L\cdot\xi_L^{\dagger}).\label{hlsb2}
\end{align}
$\ahp^{\mu}$ is related to the vector fields $V^{\mu}$ in the traditional vector-field approach
\begin{align}
\xi \ahp^{\mu}=V^{\mu},\label{apv}
\end{align}
where $\xi$ is a parameter related to the redefinition of the vector field $V^{\mu}$ in the HLS approach. The chiral transformation properties of $\ahe^{\mu}$ and $\hat\alpha_\parallel^\mu$ are
\begin{align}
{\hat{\alpha}_{\perp,\parallel }}^{\mu}\rightarrow h(x)\; {\hat{\alpha}_{\perp,\parallel }}^{\mu}\; h^{\dagger}(x).
\end{align}
The gauge field strength of the HLS gauge boson is
\begin{align}
V^{\mu\nu}=\partial^{\mu}K^{\nu}-\partial^{\nu}K^{\mu}-i\left[K^{\mu},K^{\nu}\right].\label{defV}
\end{align}
The leading order chiral Lagrangian is
\begin{align}
\mathcal{L}_{\text{H},2}=F_\pi^2\la \ahe_{\mu}\ahe^{\mu}\ra+F_\sigma^2\la \ahp_{\mu}\ahp^{\mu}\ra-\frac{1}{2g^2}\la V_{\mu\nu}V^{\mu\nu}\ra+\frac{1}{4}F_\chi^2\la\hat{\chi}_+\ra,
\end{align}
where $F_\pi$, $F_\sigma$ are the relevant decay constants,  $g$ is the HLS gauge coupling constant, $a=F_\sigma^2/F_\pi^2$, and $\hat{\chi}_+$ is related to the scalar and the pseudoscalar external sources
\begin{align}
\hat\chi_{\pm}=\hat{\chi}\pm\hat{\chi}^{\dagger}\quad(\hat\chi=\xi_L\chi\xi^\dag_R).\label{hlsb3}
\end{align}

\subsection{Building blocks and transformation properties}
The leading order chiral Lagrangian only contains four kinds of building blocks, i.e. $\ahe^{\mu}$, $\ahp^{\mu}$, $V^{\mu\nu}$ and $\hat{\chi}_+$. These building blocks are not enough to construct the higher-order chiral Lagrangians. $\hat{\chi}_-$ and the following building blocks are also necessary,
\begin{align}
&\As^{\mu\nu}= D^\mu\ahe^\nu-D^\nu\ahe^\mu-i[\ahp^\mu,\ahe^\nu]-i[\ahe^\mu,\ahp^\nu],\\
&\Hs^{\mu\nu}= D^{\mu}{\ahe^{\nu}}+ D^{\nu}{\ahe^{\mu}}-i[\ahp^{\mu},\ahe^{\nu}]+i[\ahe^{\mu},\ahp^{\nu}], \\
&\Vs^{\mu\nu}= D^\mu\ahp^\nu-D^\nu\ahp^\mu-i[\ahp^\mu,\ahp^\nu]-i[\ahe^\mu,\ahe^\nu]+V^{\mu\nu},\\
&H^{\mu\nu}= D^\mu\ahp^\nu+D^\nu\ahp^\mu.\label{hlsb4}
\end{align}
These equations give symmetric (antisymmetric) indices of $D^\mu\ahe^\nu$ and $D^\mu\ahp^\nu$. In the NLO, only the antisymmetric building blocks $\As^{\mu\nu}$ and $\Vs^{\mu\nu}$ exist \cite{harada_hidden_2003}. To the higher orders, the symmetric ones, i.e. $\Hs^{\mu\nu}$ and $H^{\mu\nu}$, will also appear. In addition, there exists another form of building blocks \cite{ma_hidden_2005}. In this work,  we choose the former one.

Any building blocks $Y$ transforms to $hYh^\dag$ under the chiral rotation. The covariant derivative of these building blocks is
\begin{align}
D^\mu Y=\partial^\mu Y-i[K^\mu, Y].
\end{align}
The form of these building blocks is completely different from those in the other approaches. Generally, the relations among these building blocks are more complicated. However, in the unitary gauge, the relations are simple \cite{tanabashi_chiral_1993,tanabashi_formulations_1996,harada_hidden_2003}. Table \ref{blhbt} presents the relations between Eqs. \eqref{hlsb1}, \eqref{hlsb2}, \eqref{hlsb3}-\eqref{hlsb4} and Eq. \eqref{bb} in the unitary gauge. The transformation properties of these building blocks are also listed in Table \ref{blhbt}. With these building blocks and their covariant derivatives, the chiral Lagrangians at any order can be constructed.
\begin{table*}[!h]
\caption{\label{blhbt}Chiral dimension (Dim), relations with pseudoscalar mesonic building blocks $O$, parity ($P$), charge conjugation ($C$) and hermiticity (h.c.) of the building blocks in the HLS approach.}
\begin{center}
\begin{tabular}{cccccc}
\hline\hline
& Dim &                               O               &  $P$               &                $C$                &               h.c.                \\
\hline
$\ahe^{\mu}$            &  1  &                      $\frac{1}{2}u^{\mu}$   &   $-\ahe_{\mu}$                   &         $(\ahe^{\mu})^T$          &           $\ahe^{\mu}$            \\
$\ahp^{\mu}$            &  1  &                  &  $\ahp_{\mu}$              &         $-(\ahp^{\mu})^T$         &           $\ahp^{\mu}$            \\
$\hat\chi_{\pm}$        &  2  &                          $\chi_{\pm}$         &   $\hat\chi_{\pm} $             &       $(\hat\chi_{\pm})^T$        &       $\pm \hat\chi_{\pm}$        \\
$\Vs^{\mu\nu}$           &  2  &                   $\frac{1}{2}\fp^{\mu\nu}$      &   $\Vs_{\mu\nu}$             &        $-(\Vs^{\mu\nu})^T$        &          $ \Vs^{\mu\nu}$          \\
$\Vt^{\mu\nu}$           &  2  &    &$\Vt_{\mu\nu}$  &        $-(\Vt^{\mu\nu})^T$        &          $\Vt^{\mu\nu}$           \\
$\Hs^{\mu\nu}$           &  2  &                    $\frac{1}{2}h^{\mu\nu}$      &$-\Hs_{\mu\nu}$               &        $(\Hs^{\mu\nu})^T$         &          $\Hs^{\mu\nu}$           \\
$\As^{\mu\nu}$           &  2  &                   $-\frac{1}{2}\fm^{\mu\nu}$     &$-\As_{\mu\nu}$               &        $(\As^{\mu\nu})^T$         &          $\As^{\mu\nu}$           \\
$H^{\mu\nu}$            &  2  &                                                            &$H_{\mu\nu}$      &          $-(H^{\mu\nu})^T$         &           $H^{\mu\nu}$            \\
$\varepsilon^{\mu\nu\lambda\rho}$ &  0  &                                                  &  $-\varepsilon_{\mu\nu\lambda\rho}$             & $\varepsilon^{\mu\nu\lambda\rho}$ & $\varepsilon^{\mu\nu\lambda\rho}$ \\
\hline\hline
\end{tabular}
\end{center}
\end{table*}

Table \ref{blhbt} also gives the power counting of HLS building blocks \cite{tanabashi_chiral_1993,harada_hidden_2003}. In addition, the $g$ parameter in the leading order chiral Lagrangians is counted as $\mathcal{O}(p^1)$ \cite{harada_hidden_2003}.

\subsection{Constraint relations}

For convenience, all constraint relations are discussed in the unitary gauge. Hence, most of the conclusions in Sec. \ref{vcr} also apply, if the relations in Table \ref{blhbt} are adopted. One difference is the EOMs \cite{harada_hidden_2003}. The EOMs in the HLS approach are
\begin{align}
&D_\mu\ahe^\mu=-i(a-1)[\ahp_\mu,\ahe^\mu]-\frac{i}{4}\frac{F_\chi^2}{F_\pi^2}\bigg(\chimh-\frac{1}{N_f}\la\chimh\ra\bigg)+\mathcal{O}(p^4)\notag\\
&D_\mu\ahp^\mu=\mathcal{O}(p^4)\notag\\
&D_{\nu}V^{\nu\mu}=g^2 F_\sigma^2\ahp^\mu+\mathcal{O}(p^4).\label{eq.23}
\end{align}
For $V^{\mu\nu}$, with the definition in Eq. \eqref{defV}, a relation similar to Bianchi identity can be obtained
\begin{align}
&D^{\rho}V^{\mu\nu}+D^{\mu}V^{\nu\rho}+D^{\nu}V^{\rho\mu}=0.\label{eq.28}
\end{align}
Now, all linear relations in the HLS approach are given.

\section{Results of chiral Lagrangians with vector mesons}\label{rcl}
The leading order chiral Lagrangians with vector mesons were obtained more than 30 years ago \cite{ecker_chiral_1989,ecker_role_1989,bando_is_1985}. We will discuss the higher-order ones. However, the higher-order chiral Lagrangians contain a lot of terms. If the calculation is done by hand, some linear relations are easy to miss. Hence, all the calculations are done with the help of computer. The difficulty of the calculations is the symbolic computation. Until now, we have not found an effective and available program/software to generate all possible terms and all linear relations. Hence, we adopt a trick to change the symbolic computation into a numerical one. This paper only gives the basic idea. The details can be found in Refs. \cite{Jiang:2014via,Jiang:2016vax,Jiang:2017yda}. To avoid the symbolic problems, we assign a number to each building block and each index. These numbers are different from one another. For example, $V\to11$, $u\to12$, $\mu\to1$, $\nu\to2$ and so on. A term can be represented by two vectors, one for the building blocks, the other one for the indices, such as $V^{\mu}u_{\mu}u^{\nu}u_{\nu}\to(11,12,12,12)$ and $(1,1,2,2)$. Comparing with the symbols, it is much easier to deal with these vectors by computer. Almost all tedious calculations are done by these vectors.

\subsection{NLO}
In Appendix \ref{app.2}, Table \ref{vnlo}  lists the chiral Lagrangians with vector mesons at the NLO in the vector-field approach. In the two-flavor case, there are a total of 81 terms, including 61 normal terms and 20 anomalous terms. In the three-flavor case, there are a total of 185 terms, including 116 normal terms and 69 anomalous terms.

For the tensor-field approach, the results in the large-$N_C$ limit are given in Ref. \cite{cirigliano_towards_2006}. In the three-flavor case, with the linear relations in Sec. \ref{vcr}, there exists another relation,
\begin{align}
{\cal O}^V_{17} \doteq & \frac{1}{2}{\cal O}^V_{11}-{\cal O}^V_{12}-{\cal O}^V_{13}-{\cal O}^V_{15}-2{\cal O}^V_{16},\label{tnlo1}
\end{align}
where ${\cal O}^V_{i}$ represents the $i$-th term in Table 1 in Ref. \cite{cirigliano_towards_2006}. Combining with the relations in Eq. \eqref{a.4}, there exist the other three relations,
\begin{align}
&{\cal O}^V_{19}\doteq -2{\cal O}^V_{5}+ \frac{1}{2}{\cal O}^V_{11} -{\cal O}^V_{12}-{\cal O}^V_{13}-{\cal O}^V_{15}-2{\cal O}^V_{16}+2{\cal O}^V_{18}-\frac{1}{2} {\cal O}^V_{20},\quad
{\cal O}^V_{21}\doteq 0,\quad{\cal O}^V_{22}\doteq 0.\label{tnlo2}
\end{align}
The terms in the left-hand side of Eqs. \eqref{tnlo1} and \eqref{tnlo2} are not independent and can be removed. There exist six terms in the $\mathcal{O}(1/N_C)$ order, i.e.
\begin{align}
&{\cal O}^V_{23}=i\la u^{\mu}u_{\mu}\ra\la W^{\nu\lambda}u_{\nu}u_{\lambda}\ra,\quad
{\cal O}^V_{24}=\la W^{\mu\nu}f_{+\mu\nu}\ra\la u^{\lambda}u_{\lambda}\ra,\quad
{\cal O}^V_{25}=\la W^{\mu\nu}{f_{+\mu}}^{\lambda}\ra\la u_{\nu}u_{\lambda}\ra,\notag\\
&{\cal O}^V_{26}=\la W^{\mu\nu}u_{\mu}\ra\la{f_{+\nu}}^{\lambda}u_{\lambda}\ra,\quad
{\cal O}^V_{27}=i\la\chip\ra\la W^{\mu\nu}u_{\mu}u_{\nu}\ra,\quad
{\cal O}^V_{28}=\la\chip\ra\la W^{\mu\nu}f_{+\mu\nu}\ra,
\end{align}
where the numbers follow the ones in Ref. \cite{cirigliano_towards_2006}. In the two-flavor case, only 16 terms are left. ${\cal O}^V_{3}$, ${\cal O}^V_{4}$, ${\cal O}^V_{9}$, ${\cal O}^V_{23}$, ${\cal O}^V_{24}$, ${\cal O}^V_{25}$, ${\cal O}^V_{26}$, ${\cal O}^V_{27}$ need to be removed.

For the anomalous part, Refs. \cite{Kampf:2011ty,Roig:2013baa} have given the whole results. With the linear relations in Sec. \ref{vcr} and Eq. \eqref{a.4}, there exist the other three relations among them, i.e.
\begin{align}
{\cal O}^{V,a}_{2}\doteq-{\cal O}^{V,a}_{1}-{\cal O}^{V,a}_{5}+{\cal O}^{V,a}_{7}+{\cal O}^{V,a}_{8},\quad {\cal O}^{V,a}_{4}\doteq-\frac{1}{2} {\cal O}^{V,a}_{15},\quad {\cal O}^{V,a}_{16}\doteq-\frac{1}{2} {\cal O}^{V,a}_{11}+\frac{1}{2}{\cal O}^{V,a}_{12}.
\end{align}
where ${\cal O}^{V,a}_{i}$ represents the $i$-th term in Eq. (25) in Ref. \cite{Kampf:2011ty}.

For the HLS approach, we repeat the results in Refs. \cite{tanabashi_chiral_1993,harada_hidden_2003}, but there also exists a correction. For the three-flavor normal terms, with Cayley-Hamilton relation \eqref{eq.24}, $y_{17}$ term in Ref. \cite{harada_hidden_2003} is relevant to the other terms,
\begin{align}
Y_{17}=&2Y_6+2Y_7+Y_8+Y_{15}+Y_{16},
\end{align}
where $Y_i$ is the term related to $y_i$. The reason is that Eq. \eqref{eq.24} is more general. Refs. \cite{tanabashi_chiral_1993,harada_hidden_2003} only adopt a special form.

For the three-flavor anomalous terms, some literature has given a similar result with different building blocks \cite{fujiwara_non-abelian_1985,Furui:1986ep,Jain:1987sz,bando_nonlinear_1988,harada_hidden_2003}. With these different building blocks, the results are
\begin{align}
\mathcal{L}_{\text{H},A,4}=&\frac{N_c}{16\pi^2}\big[
c_1\; i \varepsilon_{\mu \nu \rho \sigma} \la{\hat{\alpha}_L}^{\mu} {\hat{\alpha}_L}^{\nu} {\hat{\alpha}_L}^{\rho} {\hat{\alpha}_R}^{\sigma}-{\hat{\alpha}_R}^{\mu} {\hat{\alpha}_R}^{\nu} {\hat{\alpha}_R}^{\rho} {\hat{\alpha}_L}^{\sigma}\ra
+c_2\; i \varepsilon_{\mu \nu \rho \sigma} \la{\hat{\alpha}_L}^{\mu} {\hat{\alpha}_R}^{\nu} {\hat{\alpha}_L}^{\rho} {\hat{\alpha}_R}^{\sigma}\ra\notag\\
&\quad+c_3\;\varepsilon_{\mu \nu \rho \sigma} \la {F_V}^{\mu \nu} ({\hat{\alpha}_L}^{\rho} {\hat{\alpha}_R}^{\sigma}-{\hat{\alpha}_R}^{\rho} {\hat{\alpha}_L}^{\sigma})\ra
+ \frac{c_4}{2}\varepsilon_{\mu \nu \rho \sigma} \la {\hat{F}_L}^{\mu \nu} ({\hat{\alpha}_L}^{\rho} {\hat{\alpha}_R}^{\sigma}-{\hat{\alpha}_R}^{\rho} {\hat{\alpha}_L}^{\sigma}) \notag\\ &-{\hat{F}_R}^{\mu \nu} ({\hat{\alpha}_R}^{\rho} {\hat{\alpha}_L}^{\sigma}-{\hat{\alpha}_L}^{\rho} {\hat{\alpha}_R}^{\sigma})\ra\big].
\end{align}
With the building blocks in this paper, the results are
\begin{align}\label{hlsano}
\mathcal{L}_{\text{H},A,4}=&\;
{\cal B}_1\, i \varepsilon^{\mu \nu \lambda \rho}\langle\hat{\alpha}_{\perp \mu} \hat{\alpha}_{\perp \nu} \hat{\alpha}_{\perp \lambda} \hat{\alpha}_{\parallel \rho}\rangle
+{\cal B}_2\,\varepsilon^{\mu \nu \lambda \rho}(\langle\hat{\alpha}_{\perp \mu} \hat{\alpha}_{\parallel \nu} \Vs_{\lambda \rho}\rangle+\langle\hat{\alpha}_{\perp \mu} \Vs_{\lambda \rho}\hat{\alpha}_{\parallel \nu} \rangle)\notag\\
&+{\cal B}_3\,\varepsilon^{\mu \nu \lambda \rho}(\langle\hat{\alpha}_{\perp \mu} \hat{\alpha}_{\parallel \nu} V_{\lambda \rho}\rangle+\langle\hat{\alpha}_{\perp \mu} V_{\lambda \rho}\hat{\alpha}_{\parallel \nu} \rangle)
+{\cal B}_4\, i \varepsilon^{\mu \nu \lambda \rho}\langle\hat{\alpha}_{\perp \mu} \hat{\alpha}_{\parallel \nu} \hat{\alpha}_{\parallel \lambda} \hat{\alpha}_{\parallel \rho}\rangle.
\end{align}
These two sets of building blocks have the following relations
\begin{align}
&{\hat{\alpha}_L}^{\mu}={\hat{\alpha}_\parallel}^{\mu}-{\hat{\alpha}_\perp}^{\mu},\quad {\hat{\alpha}_R}^{\mu} ={\hat{\alpha}_\parallel}^{\mu}+{\hat{\alpha}_\perp}^{\mu},\quad \varepsilon_{\mu\nu\rho\sigma}{F_V}^{\mu \nu} = \frac{1}{2}\varepsilon_{\mu\nu\rho\sigma}V^{\mu \nu},\\
& \varepsilon_{\mu\nu\rho\sigma}{\hat{F}_L}^{\mu \nu} = \frac{1}{2}\varepsilon_{\mu\nu\rho\sigma}(\Vs^{\mu \nu} - \As^{\mu \nu}),\quad \varepsilon_{\mu\nu\rho\sigma}{\hat{F}_R}^{\mu \nu} = \frac{1}{2}\varepsilon_{\mu\nu\rho\sigma}(\Vs^{\mu \nu}+\As^{\mu \nu}).
\end{align}
The relations of the LECs are
\begin{align}
{\cal B}_1=\frac{N_c}{16\pi^2}(-4c_1+4c_2),\quad{\cal B}_2=-\frac{N_c}{16\pi^2}c_4,\quad {\cal B}_3=-\frac{N_c}{16\pi^2}c_3,\quad {\cal B}_4=\frac{N_c}{16\pi^2}(-4c_1-4c_2).
\end{align}

\subsection{NNLO}\label{nnlo}
In the vector-field approach, the numbers of the independent terms in the NLO are very large. Although the calculation in the NNLO has no problems, we consider the numbers of the terms are too large and do not give them. This also indicates an advantage of the tensor-field approach. It gives less term than the vector-field approach. Hence, this section only gives the results in the tensor-field and the HLS approaches.

The NNLO results are also presented in Appendix \ref{app.2}. Table \ref{tnnlo} lists the results in the tensor-field approach. For the HLS approach, a part of the chiral Lagrangians are related to the pseudoscalar mesonic chiral Lagrangians.
With the relations in Table \ref{blhbt}, this part of the Lagrangians can be obtained from Refs. \cite{bijnens_mesonic_1999,bijnens_anomalous_2002} directly. In order to save space, we do not present this part obviously. Table \ref{hnnlo} lists the NNLO chiral Lagrangians in the HLS approach. The numbers of these chiral Lagrangians are listed in Table \ref{numnn}.
\begin{table*}[!h]
\caption{\label{numnn}The numbers of the NNLO chiral Lagrangians with vector mesons in the tensor-field and the HLS approaches. For the HLS approach, the terms which are similar to those in the pseudoscalar mesonic chiral Lagrangians \cite{bijnens_mesonic_1999,bijnens_anomalous_2002} are ignored.}
\begin{center}
\begin{tabular}{ccccc}
  \hline\hline
   Flavor  & \multicolumn{2}{c}{$SU(2)$} & \multicolumn{2}{c}{$SU(3)$} \\
  \hline
  Approach & Tensor &        HLS         & Tensor &        HLS         \\
   Normal  &  396   &        228         &  846   &        495         \\
  Anomaly  &  177   &        173         &  676   &        251         \\
   Total   &  573   &        401         &  1522  &        746         \\
  \hline\hline
\end{tabular}
\end{center}
\end{table*}

The final results do not contain the vector singlet. If the singlet is introduced by extending to $U(3)$ symmetry, neither $\la V^{\mu}\ra$ nor $\la V^{\mu\nu}\ra$ is equal to zero. The terms contain these traces should be considered in the Lagrangians. In the vector-field and the tensor-field approaches, the results implicit a part of singlet Lagrangians, if Eq. \eqref{V9} is considered. The extra singlet Lagrangians could be obtained from the octet Lagrangians by replacing $\la VA\ra\la B\ra\to\la V\ra\la A\ra\la B\ra$, where $A$ and $B$ are some possible operators. $\la B\ra$ part could disappear. In some cases, $\la A\ra=0$ or the antisymmetric indices lead to $\la V\ra\la A\ra\la B\ra=0$. These terms should vanish. Some new Cayley-Hamilton relations could appear (Eqs. \eqref{eq.240} and \eqref{eq.24}), and some of these terms should be removed. If the singlet $\omega$ is introduced separately, $\omega$ should be considered as a build block and should be contained in the Lagrangian obviously. Only the replacement $V\to\omega$ are need. Similar to the above method, after this replacement, some terms are zeros and some new Cayley-Hamilton relations should remove some terms. However, in the HLS approach, a term contains more than one $\la\ahe^{\mu}\ra$ should be considered, there are a lot of these terms. It is beyond this work.

\section{Relations with pseudoscalar mesonic chiral Lagrangian at tree level}\label{rh}
If the vector mesonic chiral perturbation theory is regarded as a higher-energy theory, one can integrate the vector fields to obtain the pseudoscalar mesonic chiral Lagrangians. Ref. \cite{Kampf:2006bn} has given the results at the NNLO in the tensor-field approach. Ref. \cite{kampf_different_2007} has studied the problem by the first order formalism. This section discusses the HLS Lagrangian at tree level. A part of the calculations are done by Cadabra \cite{Peeters:2007wn,Peeters:2018dyg}.

Ref. \cite{harada_wilsonian_2001} has obtained the relations between the NLO pseudoscalar mesonic LECs and the HLS LECs at tree level. The method can be extended to the NNLO without any change. Integrating out the vector mesons in the HLS Lagrangian seems a little bit complicated. An equivalent method is to substitute the EOM for the vector mesons into the Lagrangian. Since this work needs to obtain the pseudoscalar mesonic chiral Lagrangians at the NNLO, the EOM for the vector mesons only needs to be expanded to the leading order. In other words, Eq. \eqref{eq.23} is enough. The other relations without vector mesons are already given in Table \ref{blhbt}. Substituting these relations into the HLS Lagrangian, the relations between the pseudoscalar mesonic and HLS LECs can be obtained. In the calculation, not all of the leading order or the NLO HLS chiral Lagrangians have an impact on the NNLO pseudoscalar mesonic chiral Lagrangians, only the following terms need to be considered
\begin{align}
&{{F_{\sigma}}}^{2} \la \ahp^{\mu} \ahp^{\mu}\ra - \frac{1}{2{g}^{2}} \la V_{\mu \nu} V^{\mu \nu}\ra+w_{5} \frac{F_{\chi}^2 }{F_{\pi}^2} \la(\ahp_{\mu} \ahe^{\mu}-\ahe_{\mu} \ahp^{\mu}) \hat{\chi}_{-}\ra+i z_{4} \la V_{\mu \nu} \ahe^{\mu} \ahe^{\nu}\ra\notag\\
&-i z_{8} \la\As_{\mu \nu} (\ahe^{\mu} \ahp^{\nu}+\ahp^{\mu} \ahe^{\nu})\ra,
\end{align}
where $F_{\chi}=F_{\pi}$ at tree level \cite{harada_wilsonian_2001,harada_hidden_2003}. Although there are many terms in Talbe \ref{hnnlo}, only a few terms which contain $V^{\mu\nu}$ have an impact on the final results.

The final results are presented in Appendix \ref{app.5}, where $\bar{C}_{i}$ represents the $i$-th LECs in the $SU(n)$ column in
Table 2 in Ref. \cite{bijnens_mesonic_1999}. ${\cal D}_{i}$ represents the $i$-th LECs in the $n$ column in Table \ref{hnnlo}. $K^W_i$ and $c^W_i$ have the same meanings as those in Ref. \cite{bijnens_anomalous_2002}.
As in Sec. \ref{nnlo}, Appendix \ref{app.5} does not contain the LECs from the terms similar to those in the pseudoscalar mesonic chiral Lagrangians. These relations are easily obtained by the relations of the building blocks in Table \ref{blhbt}. Although the relations in Appendix \ref{app.5} are in $SU(n)$ group, one can use Cayley-Hamilton relations in Refs. \cite{bijnens_mesonic_1999,Haefeli:2007ty,bijnens_anomalous_2002} to change the results in $SU(3)$ and $SU(2)$ groups respectively. Not all the LECs are listed in Appendix \ref{app.5}. The missing ones are zeros (ignore the terms similar to the pseudoscalar mesonic chiral Lagrangian). It indicates that the vector mesons only contribute to a limited part of the pseudoscalar mesonic LECs at tree level. Of course, this phenomenon has already arisen at the NLO \cite{harada_hidden_2003}.

\section{Relations between the HLS and the tensor-field approaches}\label{sec:equivalence_of_different_approachs}
Although there exist several approaches to introduce vector mesons in chiral perturbation theory, they are not irrelevant. Section \ref{rh} has integrated the vector mesons. The weak equivalence is easily found at the NNLO. As the discussion in the introduction, the relation between the vector-field and the tensor-field approaches has been studied. This section will give a simple discussion about the relation between the HLS and the tensor-field approaches. The strict relations are related to $n$-vector-meson vertices, but we have only discussed one-vector-meson vertices in the tensor-field approach. Hence, the conclusion is not complete.

The relations between the HLS approach and the tensor-field approach have been obtained through an auxiliary field method \cite{tanabashi_formulations_1996}. The idea is that an additional term
\begin{align}
\frac{1}{2}\kappa^2\la(V_\mu-a_{\parallel\mu}-\frac{1}{\kappa}D^\nu W_{\nu\mu})(V^\mu-a_{\parallel}^{\mu}-\frac{1}{\kappa}D_\lambda W^{\lambda\mu})\ra\label{aterm}
\end{align}
is added to the chiral Lagrangian in the tensor-field approach, then $W^{\mu\nu}$ field is integrated to obtain the Lagrangian in the HLS approach. $\kappa$ is just an arbitrary parameter. Operators corresponding to four-point vertices are ignored in Ref. \cite{tanabashi_chiral_1993}. When the four-point vertices are considered, the complete results are
\begin{align}
\begin{aligned}
&a =\frac{{\kappa}^{2}}{2F_\pi^2},g=\frac{\sqrt{2} M_V}{\kappa},z_1=-\frac{(\kappa-\sqrt{2}F_V)^2}{4M_V^2},z_3=\frac{\kappa(\kappa-\sqrt{2}{F_V})}{2M_V^2},z_4=\frac{ \kappa(\kappa-2 \sqrt{2} {G_V})}{M_V^2},\\
&z_5=-\frac{ {\kappa}^{2}}{M_V^2},z_6=-\frac{(\kappa-\sqrt{2}{F_V} ) (\kappa-2\sqrt{2} {G_V})}{M_V^2},z_7=\frac{\kappa(\kappa-\sqrt{2} {F_V} )}{M_V^2},\\
&y_1=-y_2=-\frac{(\kappa-2\sqrt{2} {G_V})^{2}}{2M_V^2},y_3=-y_4=- \frac{ {\kappa}^{2}}{2M_V^2},y_6=-y_7=-\frac{ \kappa(\kappa-2 \sqrt{2} {G_V})}{M_V^2},
\end{aligned}\label{lecrh}
\end{align}
where $F_V$ and $G_V$ are the leading order LECs in the tensor-field approach,  $z_i$ and $y_i$ are the NLO LECs in the HLS approach. The results related to $y_i (i=1,\cdots,7)$ are not presented in Ref. \cite{tanabashi_formulations_1996}. Eq. \eqref{lecrh} manifests that the terms at the leading order in the tensor-field approach are related to a part of the NLO terms in the HLS approach. However, if the method is extended to the NLO in the tensor-field approach, no more NLO terms are obtained in the HLS approach. There are two reasons for this problem. The first one is that the auxiliary-field term \eqref{aterm} is not unique. Only a quadratic term with vector sources is considered. To the higher order, the higher-order terms with vector source  are needed. These terms could also contain the other external sources. However, these terms are hard to be integrated. The second reason is that only one-vector vertices are considered in the tensor-field approach, but the HLS Lagrangian contains $n$-vector vertices. Although some NLO terms can be obtained in Eq. \eqref{lecrh}, these terms are related to the leading order LECs in the tensor-field approach. They are obviously not the independent terms at the NLO. In other words, if the tensor-field and the HLS approach are strong equivalence, the numbers of LECs in these two approaches need to be equal, but Eq. \eqref{lecrh} does not manifest it at the NLO. Of course, with the auxiliary field method, all terms in the tensor-field approach have corresponding terms in the HLS approach.

Some terms in the HLS approach are not obtained in Eq. \eqref{lecrh}. We will discuss which terms are related to the following terms,
\begin{align}
\la \ahp_{\mu}\ahe^{\mu}\ra\la\ahp_{\nu}\ahe^{\nu}\ra,\quad \la \ahp_{\mu}\ahp_{\nu}\ra\la\ahe^{\mu}\ahe^{\nu}\ra,\quad\la(\ahp_{\mu}\ahe^{\mu}-\ahe_{\mu}\ahp^{\mu})\hat{\chi}_-\ra.
\end{align}
Their LECs are $y_{14}$, $y_{15}$ and $w_5$, respectively. The method is very similar to that in Ref. \cite{tanabashi_formulations_1996}, but the derivation is from the HLS method to the tensor-field method. To simplify the calculation, we only consider the Lagrangian includes the kinetic term and some relevant terms, the other terms which are irrelevant to this problem are ignored. The Lagrangian is
\begin{align}
\mathcal{L}_{\text{HLS}}=&F_\sigma^2 \la \hat{\alpha}_{\parallel\mu} \ahp^{\mu}\ra-\frac{1}{2g^2} \la V_{\mu \nu} V^{\mu \nu}\ra+y_{14}\la \hat{\alpha}_{\parallel\mu}\ahe^{\mu}\ra\la\hat{\alpha}_{\parallel\nu}\ahe^{\nu}\ra+y_{15}\la \hat{\alpha}_{\parallel\mu}\hat{\alpha}_{\parallel\nu}\ra\la\ahe^{\mu}\ahe^{\nu}\ra\notag\\
&+w_{5}\la(\hat{\alpha}_{\parallel\mu}\ahe^{\mu}-\ahe^{\mu}\hat{\alpha}_{\parallel\mu})\hat{\chim}\ra+\cdots,
\end{align}
where ``$\cdots$'' denotes the irrelevant terms. Introducing an auxiliary field $W^{\mu\nu}$ into the above Lagrangian, the result is
\begin{align}
\mathcal{L'}_{\text{HLS}} =\mathcal{L}_{\text{HLS}}+ \frac{1}{2g^2 t^2} \la (W_{\mu \nu}+t V_{\mu \nu}) (W^{\mu \nu}+t V^{\mu \nu})),
\end{align}
where $t$ is an artificial parameter, and the dynamics of the system is not changed. After integrating the vector field which is implicit in $K^{\mu}$ in the HLS approach, the result is
\begin{align}
\mathcal{L}'_{\text{T}}=& -\frac{1}{g^4 t^2 F_\sigma^2}\la\nabla_{\rho} W^{\rho \mu}\nabla^{\sigma} W_{\sigma \mu}\ra+\frac{1}{2g^2t^2}\la W_{\mu\nu}W^{\mu\nu}\ra+\frac{1}{2g^2t}\la W_{\mu\nu}f_+^{\mu\nu}\ra+\frac{1}{4g^2t}i\la W_{\mu\nu}\comm*{u^{\mu}}{u^{\nu}}\ra \notag\\
&-\frac{w_{5}}{2g^2t F_\sigma^2}\la W_{\mu\nu}\comm*{u^{\mu}}{\nabla^{\nu}\chim}\ra-\frac{w_{5}}{4g^2t F_\sigma^2}\la W_{\mu\nu}\comm*{f_-^{\mu\nu}}{\chim}\ra +\frac{y_{15}}{4g^4 t^2 F_\sigma^4}\la u_{\mu} u_{\nu}\ra\la\nabla_{\rho} W^{\rho \mu}\nabla_{\sigma} W^{\sigma \nu}\ra\notag\\
&+\frac{y_{14}}{4g^4 t^2 F_\sigma^4}\la u_{\mu}\nabla_{\rho} W^{\rho \mu} \ra\la u_{\nu}\nabla_{\sigma} W^{\sigma \nu}\ra+\frac{2}{g^8 t^3 F_\sigma^4}i\la\nabla_{\rho} W^{\rho \mu} W_{\mu\nu}\nabla_{\sigma} W^{\sigma \nu}\ra\notag\\
&+\frac{w_{5}}{2g^4 t^2 F_\sigma^4}\la\nabla_{\rho}W^{\rho\mu}[W_{\mu\nu},u^{\nu}\chim-\chim u^{\nu}]\ra +\cdots,\label{astcls}
\end{align}
where "$\cdots$" represents the pseudoscalar terms and the higher-order terms, which are not relevant to this discussion. The first six terms correspond to the following Lagrangian in the tensor-field approach \cite{ecker_chiral_1989,cirigliano_towards_2006}
\begin{align}
\mathcal{L}_{\text{T}}=&-\frac{1}{2}\la\nabla_{\rho} W^{\rho \mu}\nabla^{\sigma} W_{\sigma \mu}\ra+\frac{M_V^2}{4}\la W_{\mu\nu}W^{\mu\nu}\ra+\frac{1}{2\sqrt{2}}F_V\la W_{\mu\nu}f_+^{\mu\nu}\ra+\frac{1}{2\sqrt{2}}G_Vi\la W_{\mu\nu}\comm*{u^{\mu}}{u^{\nu}}\ra\notag\\
&+D_{10}\la W_{\mu\nu}\comm*{u^{\mu}}{\nabla^{\nu}\chim}\ra+D_{20}\la W_{\mu\nu}\comm*{f_-^{\mu\nu}}{\chim}\ra,\label{lt}
\end{align}
where $D_{10}$ and $D_{20}$ are the $i$-th LECs in the NLO Lagrangians in the tensor-field approach (Table 1 in Ref. \cite{cirigliano_towards_2006}).
Comparing Eqs. \eqref{astcls} and \eqref{lt}, the following relations are obtained
\begin{align}
t=\frac{\sqrt{2}}{g^2 F_\sigma},\quad M_V=gF_\sigma,\quad F_V=F_\sigma,\quad G_V=\frac{F_\sigma}{2}.
\end{align}
These relations are consistent with the results in Refs. \cite{ecker_chiral_1989,tanabashi_formulations_1996}. KSRF I relation $F_V=2G_V$ can also be obtained from them. In Eq. \eqref{astcls}, $y_{14}$ and $y_{15}$ are related to $\la u_{\mu}\nabla_{\rho} W^{\rho \mu} \ra\la u_{\nu}\nabla_{\sigma} W^{\sigma \nu}\ra$ and $\la u_{\mu} u_{\nu}\ra\la\nabla_{\rho} W^{\rho \mu}\nabla_{\sigma} W^{\sigma \nu}\ra$, respectively. It indicates that $y_{14}$ and $y_{15}$ terms are related to two-vector vertices. $w_5$ is related to $\la W_{\mu\nu}\comm*{u^{\mu}}{\nabla^{\nu}\chim}\ra$, $\la W_{\mu\nu}\comm*{f_-^{\mu\nu}}{\chim}\ra$ and $\la\nabla_{\rho}W^{\rho\mu}[W_{\mu\nu},u^{\nu}\chim-\chim u^{\nu}]\ra$. The last term also contains a two-vector vertex. $w_5$ term combines one- and two- vector terms together. This is a general conclusion. In other words, one term in the HLS approach could be related to more than one term in the tensor-field approach. These terms could have different numbers of vector fields. This simple example indicates that some terms in the HLS approach are related to $n$-vector vertices terms in the tensor-field approach. This conclusion can be seen directly from the HLS Lagrangian too, since one $\hat{a}_\parallel$ contains one vector field. Hence, the whole equivalence needs to consider all number of vector vertices, but it is very hard to do it at present.

\section{Conclusions}\label{sec:conclusions_and_discussions}
In this work, the chiral Lagrangians with vector mesons are constructed at the NLO in the vector-field approach. The Lagrangian is also constructed at the NNLO in the tensor-field and the HLS approaches. Both the normal and the anomalous parts are considered. Some extra constraint conditions are also obtained in the tensor-field and the HLS approaches at the NLO. With these new constraint conditions, some terms in literature are linearly dependent and need to be removed. After integrating the vector field, the correspondences between the NNLO LECs of the pseudoscalar mesonic Lagrangian and the LECs in the HLS approach are obtained at tree level. Finally, the equivalent relations between the HLS and the tensor-field approaches are discussed. The whole equivalence needs consider all numbers of vector vertices. In this work, we only consider one-vector vertices in the tensor approach. The whole discussion is complicated and will be studied in the future.

Although the high-order chiral Lagrangian contains many terms, for a special problem, not all of them are needed. If one uses vector mesonic chiral Lagrangian to study pseudoscalar mesons, very few terms are relevant, such as the discussion in Sec. \ref{sec:equivalence_of_different_approachs}. At present, some complex computations can also be done by using the computer algebra system. Some tedious mathematical calculations are easy to deal with. For example, a lot of tedious calculations in Sec. \ref{rh} are done by Cadabra. This example obtains the contributions of vector mesons to the mesonic LECs by HLS approach. Only a few terms are concerned. Analogously, the high-order contributions of the properties of vector mesons could also be studied, such as the masses, electromagnetic properties and scattering processes with pseudoscalar meson of vector mesons, and so on.

\section*{acknowledgements}

We thank Professor Yong-Liang Ma for helpful discussions. Guo thanks Kasper Peeters for helpful discussions about how to use Cadabra. This work was supported by the National Science Foundation of China (NSFC) under Grant NO. 11565004, Guangxi Science Foundation under Grants No. 2018GXNSFAA281180 and No. 2017AD22006.

\appendix
\section{Some chiral Lagrangians with vector meson at the NLO and NNLO order with different approaches}\label{app.2}
\begin{longtable}{lcclcc}
\caption{The NLO chiral Lagrangians with vector mesons in the vector-field approach.}\label{vnlo}\\
\hline\hline $P_n$ & $SU(2)$ & $SU(3)$ & $P_n$ & $SU(2)$ & $SU(3)$\\
\hline\endfirsthead

\hline\hline $P_n$ & $SU(2)$ & $SU(3)$ & $P_n$ & $SU(2)$ & $SU(3)$\\
\hline\endhead

\hline\hline 
\endfoot

\hline\endlastfoot

$i\la V^{\mu}u_{\mu}u^{\nu}u^{\lambda}f_{-\nu\lambda}\ra+\mathrm{H.c.}$ & 1 & 1 & $\la\nabla^{\mu}\chi_{+}\ra\la V^{\nu}f_{+\mu\nu}\ra$ & 53 & 94  \\
$i\la V^{\mu}u^{\nu}u_{\mu}u^{\lambda}f_{-\nu\lambda}\ra+\mathrm{H.c.}$ & 2 & 2 & $i\la V^{\mu}\chip\nabla_{\mu}\chi_{+}\ra+\mathrm{H.c.}$ & 54 & 95  \\
$i\la V^{\mu}u^{\nu}u_{\nu}u^{\lambda}f_{-\mu\lambda}\ra+\mathrm{H.c.}$ & 3 & 3 & $\la V^{\mu}u_{\mu}u^{\nu}u_{\nu}\chim\ra+\mathrm{H.c.}$ & 55 & 96  \\
$i\la V^{\mu}u^{\nu}u^{\lambda}u_{\mu}f_{-\nu\lambda}\ra+\mathrm{H.c.}$ &  & 4 & $\la V^{\mu}u^{\nu}u_{\mu}u_{\nu}\chim\ra+\mathrm{H.c.}$ &  & 97  \\
$i\la V^{\mu}u^{\nu}u^{\lambda}u_{\nu}f_{-\mu\lambda}\ra+\mathrm{H.c.}$ & 4 & 5 & $\la V^{\mu}u^{\nu}u_{\nu}u_{\mu}\chim\ra+\mathrm{H.c.}$ &  & 98  \\
$i\la V^{\mu}u^{\nu}u^{\lambda}u_{\lambda}f_{-\mu\nu}\ra+\mathrm{H.c.}$ &  & 6 & $\la V^{\mu}u_{\mu}u^{\nu}\chim u_{\nu}\ra+\mathrm{H.c.}$ &  & 99  \\
$i\la V^{\mu}u_{\mu}u^{\nu}u^{\lambda}h_{\nu\lambda}\ra+\mathrm{H.c.}$ & 5 & 7 & $\la V^{\mu}u^{\nu}u_{\mu}\chim u_{\nu}\ra+\mathrm{H.c.}$ &  & 100  \\
$i\la V^{\mu}u^{\nu}u_{\mu}u^{\lambda}h_{\nu\lambda}\ra+\mathrm{H.c.}$ & 6 & 8 & $\la u^{\mu}\chim\ra\la V^{\nu}u_{\nu}u_{\mu}\ra+\mathrm{H.c.}$ &  & 101  \\
\hline
\end{longtable}

\begin{longtable}{lcclcc}
\caption{The NNLO chiral Lagrangians with vector meson in tensor-field approach}\label{tnnlo}\\
\hline\hline $P_n$ & $SU(2)$ & $SU(3)$ & $P_n$ & $SU(2)$ & $SU(3)$\\
\hline\endfirsthead

\hline\hline $P_n$ & $SU(2)$ & $SU(3)$ & $P_n$ & $SU(2)$ & $SU(3)$\\
\hline\endhead

\hline\hline 
\endfoot

\hline\endlastfoot

$i\la W^{\mu\nu}u_{\mu}u_{\nu}u^{\lambda}u_{\lambda}u^{\rho}u_{\rho}\ra+\mathrm{H.c.}$ & 1 & 1 & $\la W^{\mu\nu}{f_{-\mu}}^{\lambda}\ra\la u_{\nu}u_{\lambda}\chim\ra+\mathrm{H.c.}$ &  & 769  \\
$i\la W^{\mu\nu}u_{\mu}u_{\nu}u^{\lambda}u^{\rho}u_{\lambda}u_{\rho}\ra+\mathrm{H.c.}$ & 2 & 2 & $\la W^{\mu\nu}\nabla^{\lambda}\chi_{-}\ra\la u_{\mu}u_{\nu}u_{\lambda}\ra$ &  & 770  \\
$i\la W^{\mu\nu}u_{\mu}u_{\nu}u^{\lambda}u^{\rho}u_{\rho}u_{\lambda}\ra+\mathrm{H.c.}$ &  & 3 & $\la W^{\mu\nu}\chim\ra\la u_{\mu}u^{\lambda}f_{-\nu\lambda}\ra+\mathrm{H.c.}$ &  & 771  \\
$i\la W^{\mu\nu}u_{\mu}u^{\lambda}u_{\nu}u_{\lambda}u^{\rho}u_{\rho}\ra+\mathrm{H.c.}$ & 3 & 4 & $\la W^{\mu\nu}\nabla_{\mu}{f_{-\nu}}^{\lambda}\nabla_{\lambda}\chi_{-}\ra+\mathrm{H.c.}$ & 349 & 772  \\
$i\la W^{\mu\nu}u_{\mu}u^{\lambda}u_{\nu}u^{\rho}u_{\lambda}u_{\rho}\ra+\mathrm{H.c.}$ & 4 & 5 & $\la W^{\mu\nu}\nabla^{\lambda}f_{-\mu\nu}\nabla_{\lambda}\chi_{-}\ra+\mathrm{H.c.}$ & 350 & 773  \\
$i\la W^{\mu\nu}u_{\mu}u^{\lambda}u_{\lambda}u_{\nu}u^{\rho}u_{\rho}\ra+\mathrm{H.c.}$ &  & 6 & $\la W^{\mu\nu}\nabla^{\lambda}f_{-\mu\lambda}\nabla_{\nu}\chi_{-}\ra+\mathrm{H.c.}$ & 351 & 774  \\
$i\la W^{\mu\nu}u_{\mu}u^{\lambda}u_{\nu}u^{\rho}u_{\rho}u_{\lambda}\ra+\mathrm{H.c.}$ &  & 7 & $\la W^{\mu\nu}\nabla_{\mu}{h_{\nu}}^{\lambda}\nabla_{\lambda}\chi_{-}\ra+\mathrm{H.c.}$ & 352 & 775  \\
$i\la W^{\mu\nu}u_{\mu}u^{\lambda}u_{\lambda}u^{\rho}u_{\nu}u_{\rho}\ra+\mathrm{H.c.}$ &  & 8 & $\la W^{\mu\nu}\nabla_{\mu}\nabla^{\lambda}f_{-\nu\lambda}\chim\ra+\mathrm{H.c.}$ & 353 & 776  \\
\hline
\end{longtable}

\begin{longtable}{@{}l@{}c@{ }c@{ }c@{ }l@{}c@{ }c@{ }c}
\caption{The NNLO chiral Lagrangians with vector mesons in the HLS approach. Some terms similar to Refs. \cite{bijnens_mesonic_1999,bijnens_anomalous_2002} are ignored. These terms can be obtained by the relations in Table \ref{blhbt} directly. $n$ is a total serial number.}\label{hnnlo}\\
\hline\hline $P_n$ & $n$ & $SU(2)$ & $SU(3)$ & $P_n$ & $n$ & $SU(2)$ & $SU(3)$\\
\hline\endfirsthead

\hline\hline $P_n$ & $n$ & $SU(2)$ & $SU(3)$ & $P_n$ & $n$ & $SU(2)$ & $SU(3)$\\
\hline\endhead

\hline\hline 
\endfoot

\hline\endlastfoot

$i\la{\hat{a}_{\perp}}^{\mu}\hat{a}_{\perp\mu}{\hat{a}_{\perp}}^{\nu}{\hat{a}_{\perp}}^{\lambda}V_{\nu\lambda}\ra+\mathrm{H.c.}$ & 1 & 1 & 1 & $i\la{\hat{a}_{\perp}}^{\mu}{\hat{a}_{\parallel}}^{\nu}\hat{a}_{\parallel\nu}{\Hs_{\mu}}^{\lambda}\hat{a}_{\parallel\lambda}\ra+\mathrm{H.c.}$ & 387 &  & 387  \\
$i\la{\hat{a}_{\perp}}^{\mu}{\hat{a}_{\perp}}^{\nu}\hat{a}_{\perp\mu}{\hat{a}_{\perp}}^{\lambda}V_{\nu\lambda}\ra+\mathrm{H.c.}$ & 2 & 2 & 2 & $i\la{\hat{a}_{\perp}}^{\mu}{\hat{a}_{\parallel}}^{\nu}{\hat{a}_{\parallel}}^{\lambda}\Hs_{\mu\nu}\hat{a}_{\parallel\lambda}\ra+\mathrm{H.c.}$ & 388 & 190 & 388  \\
$i\la{\hat{a}_{\perp}}^{\mu}{\hat{a}_{\perp}}^{\nu}\hat{a}_{\perp\nu}{\hat{a}_{\perp}}^{\lambda}V_{\mu\lambda}\ra$ & 3 &  & 3 & $i\la{\hat{a}_{\perp}}^{\mu}{\hat{a}_{\parallel}}^{\nu}{\hat{a}_{\parallel}}^{\lambda}\Hs_{\mu\lambda}\hat{a}_{\parallel\nu}\ra+\mathrm{H.c.}$ & 389 &  & 389  \\
$i\la{\hat{a}_{\perp}}^{\mu}{\hat{a}_{\perp}}^{\nu}{\hat{a}_{\perp}}^{\lambda}\hat{a}_{\perp\mu}V_{\nu\lambda}\ra$ & 4 &  & 4 & $i\la{\hat{a}_{\perp}}^{\mu}{\hat{a}_{\parallel}}^{\nu}{\hat{a}_{\parallel}}^{\lambda}\Hs_{\nu\lambda}\hat{a}_{\parallel\mu}\ra+\mathrm{H.c.}$ & 390 &  & 390  \\
$i\la{\hat{a}_{\perp}}^{\mu}\hat{a}_{\perp\mu}\ra\la{\hat{a}_{\perp}}^{\nu}{\hat{a}_{\perp}}^{\lambda}V_{\nu\lambda}\ra$ & 5 &  & 5 & $i\la{\hat{a}_{\perp}}^{\mu}\hat{a}_{\parallel\mu}{\hat{a}_{\parallel}}^{\nu}{\hat{a}_{\parallel}}^{\lambda}\As_{\nu\lambda}\ra+\mathrm{H.c.}$ & 391 & 191 & 391  \\
$\la{\hat{a}_{\perp}}^{\mu}\hat{a}_{\perp\mu}V^{\nu\lambda}\Vs_{\nu\lambda}\ra+\mathrm{H.c.}$ & 6 & 3 & 6 & $i\la{\hat{a}_{\perp}}^{\mu}{\hat{a}_{\parallel}}^{\nu}\hat{a}_{\parallel\mu}{\hat{a}_{\parallel}}^{\lambda}\As_{\nu\lambda}\ra+\mathrm{H.c.}$ & 392 & 192 & 392  \\
\hline
\end{longtable}

\section{Relations between the NNLO pseudoscalar mesonic LECs and the LECs in the HLS approach at tree level}\label{app.5}
These relations do not contain the LECs from the terms similar to those in the pseudoscalar mesonic chiral Lagrangians \cite{bijnens_mesonic_1999,Haefeli:2007ty}. They are easily obtained by Table \ref{blhbt}. Some LECs are zeros and they are not listed below.
\begin{align}
 \bar{C}_{1} \leftrightarrow &  - \frac{1}{16} C_{22} +\frac{1}{8} C_{23} +\frac{1}{16} C_{26} +\frac{1}{32} C_{39}  - \frac{1}{16} C_{40} +\frac{1}{32 F_\sigma^2 g^4} - \frac{1}{32 F_\sigma^2 g^2} {z_4} \\
\bar{C}_{3} \leftrightarrow & \frac{1}{32} C_{22}  - \frac{1}{32} C_{23} +\frac{1}{64} C_{25}  \\
\bar{C}_{5} \leftrightarrow &  - \frac{1}{32} C_{22} +\frac{3}{32} C_{23} +\frac{1}{64} C_{25} +\frac{1}{16} C_{26} +\frac{1}{32} C_{39}  - \frac{1}{16} C_{40} +\frac{1}{32 F_\sigma^2 g^4} - \frac{1}{32 F_\sigma^2 g^2} {z_4} \\
\bar{C}_{7} \leftrightarrow & \frac{1}{16}  C_{16} \\
\bar{C}_{8} \leftrightarrow & \frac{1}{16} C_{17} +\frac{1}{8} C_{38}  \\
\bar{C}_{11} \leftrightarrow & \frac{1}{8} C_{15} +\frac{1}{8} C_{37}  \\
\bar{C}_{13} \leftrightarrow &  - \frac{1}{8} C_{15}  - \frac{1}{16} C_{16}  - \frac{1}{8} C_{37}  \\
\bar{C}_{14} \leftrightarrow &  - \frac{1}{16} C_{17}  - \frac{1}{8} C_{38}  \\
\bar{C}_{28} \leftrightarrow &  - \frac{1}{16} C_{21} +\frac{1}{32} C_{25} +\frac{1}{64} C_{39}  - \frac{1}{32} C_{40} +\frac{1}{64 F_\sigma^2 g^4} - \frac{1}{16 F_\sigma^2 g^2} {w_5} - \frac{1}{64 F_\sigma^2 g^2} {z_4} \\
\bar{C}_{30} \leftrightarrow &  - \frac{1}{16 N_f} C_{22} +\frac{3}{16 N_f} C_{23} +\frac{1}{32 N_f} C_{25} +\frac{1}{8 N_f} C_{26}+\frac{1}{16 N_f} C_{39}  - \frac{1}{8 N_f} C_{40} \notag\\&+\frac{1}{16 F_\sigma^2 N_f g^4} - \frac{1}{16 F_\sigma^2 N_f g^2} {z_4} \\
\bar{C}_{31} \leftrightarrow & \frac{1}{8} C_{21} +\frac{1}{16} C_{22}  - \frac{3}{16} C_{23}  - \frac{3}{32} C_{25}  - \frac{1}{8} C_{26}  - \frac{3}{32} C_{39} +\frac{3}{16} C_{40}  - \frac{3}{32 F_\sigma^2 g^4}\notag\\&+\frac{1}{8 F_\sigma^2 g^2} {w_5}+\frac{3}{32 F_\sigma^2 g^2} {z_4} \\
\bar{C}_{33} \leftrightarrow &  - \frac{1}{8} C_{21}  - \frac{1}{16} C_{22} +\frac{1}{8} C_{23} +\frac{1}{16} C_{25} +\frac{1}{16} C_{26} +\frac{1}{16} C_{39}  - \frac{1}{8} C_{40} +\frac{1}{16 F_\sigma^2 g^4} \notag\\&- \frac{1}{8 F_\sigma^2 g^2} {w_5} - \frac{1}{16 F_\sigma^2 g^2} {z_4} \\
\bar{C}_{34} \leftrightarrow & \frac{1}{16 N_f} C_{22}  - \frac{3}{16 N_f} C_{23}  - \frac{1}{32 N_f} C_{25}  - \frac{1}{8 N_f} C_{26}  - \frac{1}{16 N_f} C_{39} +\frac{1}{8 N_f} C_{40} \notag\\& - \frac{1}{16 F_\sigma^2 N_f g^4}+\frac{1}{16 F_\sigma^2 N_f g^2} {z_4} \\
\bar{C}_{36} \leftrightarrow &  - \frac{1}{32 N_f^2} C_{22} +\frac{3}{32 N_f^2} C_{23} +\frac{1}{64 N_f^2} C_{25} +\frac{1}{16 N_f^2} C_{26} +\frac{1}{32 N_f^2} C_{39}  - \frac{1}{16 N_f^2} C_{40} \notag \\
&+\frac{1}{32 F_\sigma^2 N_f^2 g^4}- \frac{1}{32 F_\sigma^2 N_f^2 g^2} {z_4} \\
\bar{C}_{37} \leftrightarrow & \frac{1}{8} C_{21} +\frac{1}{32} C_{22}  - \frac{1}{32} C_{23}  - \frac{3}{64} C_{25}  - \frac{1}{32} C_{39} +\frac{1}{16} C_{40}  - \frac{1}{32 F_\sigma^2 g^4}+\frac{1}{8 F_\sigma^2 g^2} {w_5}\notag \\
&+\frac{1}{32 F_\sigma^2 g^2} {z_4} \\
\bar{C}_{49} \leftrightarrow & \frac{1}{64} C_{3} +\frac{1}{16} C_{22}  - \frac{1}{8} C_{23}  - \frac{1}{16} C_{25}  - \frac{1}{16} C_{26} +\frac{1}{64} C_{30}  - \frac{1}{32} C_{39} +\frac{1}{8} C_{40} +\frac{1}{64} C_{41}  \notag \\
&- \frac{1}{32 F_\sigma^2 g^4}+\frac{1}{32 F_\sigma^2 g^2} {z_4} \\
\bar{C}_{50} \leftrightarrow & \frac{1}{64} C_{5} +\frac{1}{32} C_{33}  \\
\bar{C}_{52} \leftrightarrow & \frac{1}{32} C_{1} +\frac{1}{16} C_{22}  - \frac{1}{16} C_{23}  - \frac{1}{32} C_{25}  - \frac{1}{16} C_{26} +\frac{1}{32} C_{28} +\frac{1}{32} C_{31}  - \frac{1}{32} C_{39} +\frac{1}{16} C_{40}  \notag\\
&- \frac{1}{32 F_\sigma^2 g^4}+\frac{1}{32 F_\sigma^2 g^2} {z_4} \\
\bar{C}_{54} \leftrightarrow &  - \frac{1}{32} C_{1} +\frac{1}{32} C_{2}  - \frac{1}{64} C_{3}  - \frac{3}{16} C_{22} +\frac{1}{4} C_{23} +\frac{3}{16} C_{25} +\frac{1}{8} C_{26}  - \frac{1}{32} C_{28} +\frac{1}{64} C_{29}  \notag \\
&- \frac{1}{32} C_{30}  - \frac{1}{16} C_{31}  +\frac{1}{16} C_{39}  - \frac{5}{16} C_{40}  - \frac{3}{64} C_{41} +\frac{1}{16 F_\sigma^2 g^4} - \frac{1}{16 F_\sigma^2 g^2} {z_4} \\
\bar{C}_{55} \leftrightarrow & \frac{1}{32} C_{34} \\
\bar{C}_{57} \leftrightarrow &  - \frac{1}{64} C_{5}  - \frac{1}{32} C_{33}  \\
\bar{C}_{58} \leftrightarrow &  - \frac{1}{64} C_{4} +\frac{1}{16} C_{23} +\frac{1}{32} C_{25} +\frac{1}{64} C_{29}  - \frac{1}{32} C_{32}  - \frac{1}{16} C_{40}  - \frac{1}{64} C_{41}  \\
\bar{C}_{60} \leftrightarrow &  - \frac{1}{32} C_{2} +\frac{1}{64} C_{4} +\frac{1}{16} C_{22}  - \frac{1}{8} C_{23}  - \frac{1}{8} C_{25}  - \frac{1}{32} C_{29} +\frac{1}{64} C_{30} +\frac{1}{32} C_{31} \notag \\
&+\frac{1}{32} C_{32} +\frac{3}{16} C_{40} +\frac{3}{64} C_{41}  \\
\bar{C}_{62} \leftrightarrow  &  - \frac{1}{32}  C_{34} \\
\bar{C}_{64} \leftrightarrow & \frac{1}{32} C_{1} +\frac{1}{16} C_{6} +\frac{1}{32} C_{9} +\frac{1}{8} C_{22}  - \frac{1}{16} C_{23}  - \frac{1}{32} C_{25}  - \frac{1}{16} C_{26} +\frac{1}{16} C_{28} +\frac{1}{16} C_{31}  \notag \\
&- \frac{1}{32} C_{39} +\frac{1}{16} C_{40} - \frac{1}{32 F_\sigma^2 g^4}+\frac{1}{32 F_\sigma^2 g^2} {z_4} \\
\bar{C}_{65} \leftrightarrow & \frac{1}{32} C_{5} +\frac{1}{16} C_{11} +\frac{1}{8} C_{33}  \\
\bar{C}_{66} \leftrightarrow & \frac{1}{32} C_{4}  - \frac{1}{16} C_{7} +\frac{1}{16} C_{10}  - \frac{1}{4} C_{21}  - \frac{1}{8} C_{22}  - \frac{1}{8} C_{23}  - \frac{1}{16} C_{25} +\frac{1}{16} C_{27}  - \frac{1}{16} C_{29} \notag \\
&+\frac{1}{8} C_{32} +\frac{1}{32} C_{36} +\frac{1}{8} C_{39} +\frac{1}{8} C_{40} +\frac{3}{32} C_{41} +\frac{1}{8 F_\sigma^2 g^4} - \frac{1}{4 F_\sigma^2 g^2} {w_5} - \frac{1}{16 F_\sigma^2 g^2} {z_4} \\
\bar{C}_{67} \leftrightarrow & \frac{1}{32} C_{3} +\frac{1}{16} C_{8}  - \frac{1}{4} C_{21} +\frac{1}{8} C_{22}  - \frac{1}{4} C_{23}  - \frac{1}{8} C_{25}  - \frac{1}{8} C_{26} +\frac{1}{16} C_{27} +\frac{1}{16} C_{30} \notag \\
&  +\frac{1}{32} C_{36} +\frac{1}{16} C_{39} +\frac{1}{4} C_{40}+\frac{3}{32} C_{41} +\frac{1}{16 F_\sigma^2 g^4} - \frac{1}{4 F_\sigma^2 g^2} {w_5} \\
\bar{C}_{68} \leftrightarrow & \frac{1}{32} C_{2} +\frac{1}{32} C_{7}  - \frac{1}{32} C_{8}  - \frac{1}{32} C_{9} +\frac{1}{4} C_{21}  - \frac{1}{16} C_{22} +\frac{1}{16} C_{23} +\frac{3}{32} C_{25}  - \frac{1}{16} C_{27}\notag \\
&  +\frac{1}{32} C_{29}  - \frac{1}{32} C_{30}  - \frac{1}{16} C_{31}  - \frac{1}{32} C_{36}  - \frac{1}{8} C_{39}  - \frac{1}{8} C_{40}  - \frac{3}{32} C_{41}  - \frac{1}{8 F_\sigma^2 g^4}\notag \\
& +\frac{1}{4 F_\sigma^2 g^2} {w_5}+\frac{1}{16 F_\sigma^2 g^2} {z_4} \\
\bar{C}_{70} \leftrightarrow & \frac{1}{32} C_{12} +\frac{1}{16} C_{34}  \\
\bar{C}_{71} \leftrightarrow & \frac{1}{8} C_{6} +\frac{1}{8} C_{21} +\frac{1}{8} C_{22}  - \frac{7}{32} C_{27} +\frac{1}{16} C_{28}  - \frac{1}{4} C_{39} +\frac{1}{2} C_{40}  - \frac{1}{4 F_\sigma^2 g^4}\notag \\
& +\frac{1}{8 F_\sigma^2 g^2} {w_5}+\frac{1}{32 F_\sigma^2 g^2} {z_4} \\
\bar{C}_{72} \leftrightarrow & \frac{1}{16} C_{11} +\frac{1}{16} C_{33}  \\
\bar{C}_{73} \leftrightarrow & \frac{1}{16} C_{10}  - \frac{1}{8} C_{21}  - \frac{1}{8} C_{22} +\frac{7}{32} C_{27} +\frac{1}{16} C_{32} +\frac{1}{4} C_{39}  - \frac{1}{2} C_{40} +\frac{1}{4 F_\sigma^2 g^4} - \frac{1}{8 F_\sigma^2 g^2} {w_5}\notag \\
&  - \frac{1}{32 F_\sigma^2 g^2} {z_4} \\
\bar{C}_{75} \leftrightarrow & \frac{1}{8} C_{7}  - \frac{1}{16} C_{14} +\frac{1}{4} C_{21} +\frac{5}{16} C_{27} +\frac{1}{16} C_{29}  - \frac{1}{8} C_{36} +\frac{1}{4} C_{39}  - \frac{5}{4} C_{40}  - \frac{3}{16} C_{41}\notag \\
&  +\frac{1}{4 F_\sigma^2 g^4}+\frac{1}{4 F_\sigma^2 g^2} {w_5}+\frac{1}{16 F_\sigma^2 g^2} {z_4} \\
\bar{C}_{76} \leftrightarrow & \frac{1}{8} C_{8} +\frac{1}{16} C_{14}  - \frac{1}{2} C_{21} +\frac{1}{4} C_{27} +\frac{1}{16} C_{30} +\frac{1}{8} C_{36} +\frac{3}{8} C_{39} +\frac{3}{16} C_{41} +\frac{3}{8 F_\sigma^2 g^4}\notag \\
&  - \frac{1}{2 F_\sigma^2 g^2} {w_5} - \frac{1}{8 F_\sigma^2 g^2} {z_4} \\
\bar{C}_{77} \leftrightarrow & \frac{1}{16} C_{12} +\frac{1}{16} C_{34}  \\
\bar{C}_{78} \leftrightarrow &  - \frac{1}{16} C_{9} +\frac{1}{8} C_{21}  - \frac{9}{32} C_{27}  - \frac{1}{16} C_{31}  - \frac{5}{16} C_{39} +\frac{5}{8} C_{40}  - \frac{5}{16 F_\sigma^2 g^4}+\frac{1}{8 F_\sigma^2 g^2} {w_5}\notag \\
& +\frac{1}{32 F_\sigma^2 g^2} {z_4} \\
\bar{C}_{80} \leftrightarrow &  - \frac{1}{16} C_{13}  - \frac{1}{16} C_{35}  \\
\bar{C}_{81} \leftrightarrow & \frac{1}{2} C_{18} +\frac{1}{4} C_{37}  \\
\bar{C}_{82} \leftrightarrow & \frac{1}{4} C_{19} +\frac{1}{4} C_{38}  \\
\bar{C}_{83} \leftrightarrow & \frac{1}{8} C_{15} +\frac{1}{4} C_{18} +\frac{1}{4} C_{37}  \\
\bar{C}_{84} \leftrightarrow & \frac{1}{8} C_{17} +\frac{1}{4} C_{19} +\frac{1}{2} C_{38}  \\
\bar{C}_{85} \leftrightarrow & \frac{1}{8} C_{16} \\
\bar{C}_{86} \leftrightarrow & \frac{1}{32} C_{25} +\frac{1}{32} C_{26} +\frac{1}{32} C_{39}  - \frac{1}{16} C_{40} +\frac{1}{32 F_\sigma^2 g^4} - \frac{1}{32 F_\sigma^2 g^2} {z_4} - \frac{1}{32 F_\sigma^2 g^2} {z_8} \\
\bar{C}_{89} \leftrightarrow &  - \frac{1}{32} C_{25}  - \frac{1}{32} C_{26}  - \frac{1}{32} C_{39} +\frac{1}{16} C_{40}  - \frac{1}{32 F_\sigma^2 g^4}+\frac{1}{32 F_\sigma^2 g^2} {z_4}+\frac{1}{32 F_\sigma^2 g^2} {z_8} \\
\bar{C}_{90} \leftrightarrow & \frac{1}{8} C_{21} +\frac{1}{16} C_{22}  - \frac{1}{8} C_{23}  - \frac{1}{16} C_{25}  - \frac{1}{16} C_{26} +\frac{3}{32} C_{27} +\frac{1}{32} C_{39}  - \frac{1}{16} C_{40} +\frac{1}{32 F_\sigma^2 g^4}\notag \\
&+\frac{1}{8 F_\sigma^2 g^2} {w_5}+\frac{1}{16 F_\sigma^2 g^2} {z_4} \\
\bar{C}_{92} \leftrightarrow &  - \frac{1}{8} C_{21}  - \frac{1}{32} C_{22} +\frac{1}{32} C_{23} +\frac{3}{64} C_{25}  - \frac{3}{32} C_{27}  - \frac{1}{16} C_{39} +\frac{1}{8} C_{40}  - \frac{1}{16 F_\sigma^2 g^4} \notag \\
&- \frac{1}{8 F_\sigma^2 g^2} {w_5} - \frac{1}{32 F_\sigma^2 g^2} {z_4} \\
\bar{C}_{94} \leftrightarrow &  - \frac{1}{4} C_{21}  - \frac{1}{16} C_{22} +\frac{1}{8} C_{23}  - \frac{1}{16} C_{24} +\frac{1}{8} C_{25}  - \frac{1}{16} C_{27}  - \frac{1}{4 F_\sigma^2 g^2} {w_5} - \frac{1}{16 F_\sigma^2 g^2} {z_4} \\
\bar{C}_{95} \leftrightarrow & \frac{1}{16} C_{22}  - \frac{1}{8} C_{23} +\frac{1}{16} C_{24}  - \frac{1}{16} C_{25}  - \frac{1}{16} C_{26}  - \frac{1}{16} C_{39} +\frac{1}{8} C_{40}  - \frac{1}{16 F_\sigma^2 g^4}\notag \\
&+\frac{1}{16 F_\sigma^2 g^2} {z_4}+\frac{1}{16 F_\sigma^2 g^2} {z_8} \\
\bar{C}_{97} \leftrightarrow & \frac{1}{8} C_{21} +\frac{1}{32} C_{22}  - \frac{3}{32} C_{23}  - \frac{3}{64} C_{25}  - \frac{1}{32} C_{26} +\frac{1}{32} C_{27} +\frac{1}{8 F_\sigma^2 g^2} {w_5}\notag \\
&+\frac{1}{32 F_\sigma^2 g^2} {z_4} - \frac{1}{32 F_\sigma^2 g^2} {z_8} \\
\bar{C}_{100} \leftrightarrow &  - \frac{1}{8} C_{21}  - \frac{1}{8} C_{22} +\frac{1}{8} C_{23} +\frac{9}{32} C_{27} +\frac{5}{16} C_{39}  - \frac{5}{8} C_{40} +\frac{5}{16 F_\sigma^2 g^4} - \frac{1}{8 F_\sigma^2 g^2} {w_5} \notag \\
&- \frac{1}{32 F_\sigma^2 g^2} {z_4} \\
\bar{C}_{101} \leftrightarrow & \frac{3}{16} C_{14} +\frac{1}{8} C_{21} +\frac{1}{16} C_{22} +\frac{1}{16} C_{24} +\frac{3}{32} C_{27} +\frac{3}{16} C_{36} +\frac{1}{16} C_{39} +\frac{5}{8} C_{40} +\frac{3}{16} C_{41}\notag \\
& +\frac{1}{16 F_\sigma^2 g^4}+\frac{1}{8 F_\sigma^2 g^2} {w_5}+\frac{1}{32 F_\sigma^2 g^2} {z_4} \\
\bar{C}_{104} \leftrightarrow &  - \frac{1}{4} C_{20}  - \frac{1}{16} C_{21}  - \frac{5}{64} C_{27}  - \frac{3}{32} C_{39} +\frac{3}{16} C_{40}  - \frac{3}{32 F_\sigma^2 g^4}+\frac{1}{16 F_\sigma^2 g^2} {w_5}\notag \\
&+\frac{1}{64 F_\sigma^2 g^2} {z_4} \\
\bar{C}_{105} \leftrightarrow &  - \frac{1}{4} C_{20}  - \frac{3}{16} C_{21}  - \frac{1}{16} C_{22} +\frac{1}{16} C_{23} +\frac{1}{16} C_{25} +\frac{1}{32} C_{26} +\frac{3}{64} C_{39}  - \frac{3}{32} C_{40} \notag \\
&+\frac{3}{64 F_\sigma^2 g^4} - \frac{1}{16 F_\sigma^2 g^2} {w_5} - \frac{3}{64 F_\sigma^2 g^2} {z_4} - \frac{1}{32 F_\sigma^2 g^2} {z_8} \\
\bar{C}_{109} \leftrightarrow & \frac{1}{8} C_{27} +\frac{1}{8} C_{39}  - \frac{1}{4} C_{40} +\frac{1}{8 F_\sigma^2 g^4} \\
\bar{C}_{110} \leftrightarrow & \frac{1}{4} C_{21}  - \frac{1}{8} C_{25}  - \frac{1}{16} C_{27}  - \frac{1}{8} C_{39} +\frac{1}{4} C_{40}  - \frac{1}{8 F_\sigma^2 g^4}+\frac{1}{4 F_\sigma^2 g^2} {w_5}+\frac{1}{16 F_\sigma^2 g^2} {z_4} \\
\bar{C}_{111} \leftrightarrow &  - \frac{1}{8} C_{22} +\frac{9}{16} C_{27} +\frac{5}{8} C_{39}  - \frac{5}{4} C_{40} +\frac{5}{8 F_\sigma^2 g^4} - \frac{1}{16 F_\sigma^2 g^2} {z_4} - \frac{1}{8 F_\sigma^2 g^2} {z_8} \\
\bar{C}_{112} \leftrightarrow & \frac{1}{4} C_{21} +\frac{1}{8} C_{22} +\frac{1}{8} C_{26} +\frac{1}{4 F_\sigma^2 g^2} {w_5} \\
\bar{C}_{114} \leftrightarrow & \frac{1}{2} C_{14} + C_{27} +\frac{1}{2} C_{36} + C_{39} +\frac{1}{2} C_{41} +{{F_\sigma}}^{-2} {g}^{-4} \\
\bar{C}_{115} \leftrightarrow &  - \frac{1}{4} C_{27}  - \frac{1}{4} C_{39} +\frac{1}{2} C_{40}  - \frac{1}{4 F_\sigma^2 g^4} \\
K_{1}^W \leftrightarrow & \frac{1}{4}{\cal D}_{496} - \frac{1}{8}{\cal D}_{497} - \frac{1}{32}{\cal D}_{502} - \frac{1}{16}{\cal D}_{503} - \frac{1}{4}{\cal D}_{518} - \frac{1}{8}{\cal D}_{520} - \frac{3}{64}{\cal D}_{521} + \frac{1}{32 F_\sigma^2 g^2} {\cal B}_{1} \notag \\
&+ \frac{1}{8 F_\sigma^2 g^2} {\cal B}_{3} \\
K_{2}^W \leftrightarrow & - \frac{1}{4} {\cal D}_{516}\\
K_{4}^W \leftrightarrow & \frac{1}{8}{\cal D}_{496} - \frac{1}{4}{\cal D}_{499} - \frac{1}{8}{\cal D}_{517} - \frac{1}{4}{\cal D}_{518}\\
K_{5}^W \leftrightarrow &  - \frac{1}{8}{\cal D}_{497} - \frac{1}{32}{\cal D}_{511}+\frac{1}{16}{\cal D}_{512}+\frac{1}{4}{\cal D}_{517} - \frac{1}{8}{\cal D}_{520} - \frac{1}{16}{\cal D}_{521} + \frac{1}{64 F_\sigma^2 g^2} {\cal B}_{1}\notag \\
& + \frac{1}{16 F_\sigma^2 g^2} {\cal B}_{2} + \frac{1}{8 F_\sigma^2 g^2} {\cal B}_{3}\\
K_{6}^W \leftrightarrow & \frac{1}{8}{\cal D}_{498} - \frac{1}{4}{\cal D}_{500} - \frac{1}{2}{\cal D}_{519} - \frac{1}{32 N_f}{\cal D}_{511}+\frac{1}{16 N_f}{\cal D}_{512} - \frac{1}{8 N_f}{\cal D}_{520} - \frac{1}{16 N_f}{\cal D}_{521} \notag\\
& + \frac{1}{16 N_f F_\sigma^2 g^2} {\cal B}_{2} + \frac{1}{16 N_f F_\sigma^2 g^2} {\cal B}_{3}  + \frac{1}{64 N_f F_\sigma^2 g^2} {\cal B}_{1} + \frac{1}{16 N_f F_\sigma^2 g^2} {\cal B}_{3}\\
K_{7}^W \leftrightarrow & \frac{1}{2}{\cal D}_{499}+\frac{1}{16}{\cal D}_{509}+\frac{1}{64}{\cal D}_{511}+\frac{1}{32}{\cal D}_{512}+\frac{1}{4}{\cal D}_{518}+\frac{1}{16}{\cal D}_{520}+\frac{1}{64}{\cal D}_{521} \notag\\
&- \frac{1}{16 F_\sigma^2 g^2} {\cal B}_{2}- \frac{1}{16 F_\sigma^2 g^2} {\cal B}_{3}\\
K_{8}^W \leftrightarrow & \frac{1}{4}{\cal D}_{500}+\frac{1}{4}{\cal D}_{519} - \frac{1}{16 N_f}{\cal D}_{509} - \frac{1}{64 N_f}{\cal D}_{511} - \frac{1}{32 N_f}{\cal D}_{512} - \frac{1}{16 N_f}{\cal D}_{520} \notag\\
&- \frac{1}{64 N_f}{\cal D}_{521} + \frac{1}{16 N_f F_\sigma^2 g^2} {\cal B}_{2}  + \frac{1}{16 N_f F_\sigma^2 g^2} {\cal B}_{3} \\
K_{11}^W \leftrightarrow & - \frac{1}{4} {\cal D}_{516} \\
K_{12}^W \leftrightarrow & \frac{1}{16}{\cal D}_{503} - \frac{1}{16}{\cal D}_{504}+\frac{1}{16}{\cal D}_{520}+\frac{1}{32}{\cal D}_{521}- \frac{1}{64 F_\sigma^2 g^2} {\cal B}_{1}- \frac{1}{16 F_\sigma^2 g^2} {\cal B}_{3}\\
K_{13}^W  \leftrightarrow & \frac{1}{16}{\cal D}_{502}+\frac{1}{8}{\cal D}_{509}+\frac{1}{32}{\cal D}_{511} - \frac{1}{16}{\cal D}_{512}+\frac{3}{16}{\cal D}_{520}+\frac{1}{16}{\cal D}_{521}+\frac{1}{32}F {\cal B}_{1}+\frac{1}{16}F {\cal B}_{2}\notag\\
&+\frac{3}{16}F {\cal B}_{3} \\
K_{14}^W  \leftrightarrow & \frac{1}{16}{\cal D}_{503}+\frac{1}{32}{\cal D}_{520}+\frac{1}{32}F {\cal B}_{2}+\frac{1}{32}F {\cal B}_{3} \\
K_{15}^W  \leftrightarrow & \frac{1}{16}{\cal D}_{504} - \frac{1}{32}{\cal D}_{511}+\frac{1}{16}{\cal D}_{512} - \frac{3}{32}{\cal D}_{520} - \frac{1}{16}{\cal D}_{521} - \frac{1}{64}F {\cal B}_{1} - \frac{1}{32}F {\cal B}_{2} - \frac{3}{32}F {\cal B}_{3} \\
K_{16}^W  \leftrightarrow &\frac{1}{16}{\cal D}_{506} - \frac{1}{16}{\cal D}_{507} - \frac{1}{16}{\cal D}_{508} - \frac{1}{16}{\cal D}_{520} - \frac{1}{32}{\cal D}_{521} - \frac{1}{64}F {\cal B}_{1} - \frac{1}{16}F {\cal B}_{3}\\
K_{17}^W \leftrightarrow &  - \frac{1}{32}{\cal D}_{507} - \frac{1}{64}{\cal D}_{521}\\
K_{19}^W \leftrightarrow &  - \frac{1}{16}{\cal D}_{505}+\frac{1}{16}{\cal D}_{506} - \frac{1}{8}{\cal D}_{510} - \frac{1}{32}{\cal D}_{511} - \frac{1}{16}{\cal D}_{512}+\frac{5}{32}{\cal D}_{520} - \frac{1}{16}{\cal D}_{521}\notag\\
&- \frac{1}{32 F_\sigma^2 g^2} {\cal B}_{1}- \frac{1}{32 F_\sigma^2 g^2} {\cal B}_{2}- \frac{5}{32 F_\sigma^2 g^2} {\cal B}_{3}\\
K_{20}^W \leftrightarrow & \frac{1}{16}{\cal D}_{506} - \frac{1}{16}{\cal D}_{507} - \frac{1}{16}{\cal D}_{511} - \frac{1}{32}{\cal D}_{520} - \frac{3}{32}{\cal D}_{521} + \frac{1}{32 F_\sigma^2 g^2} {\cal B}_{2} + \frac{1}{32 F_\sigma^2 g^2} {\cal B}_{3}\\
K_{21}^W \leftrightarrow & \frac{1}{32}{\cal D}_{506} - \frac{1}{16}{\cal D}_{507} - \frac{1}{16}{\cal D}_{508}+\frac{1}{16}{\cal D}_{512} - \frac{7}{64}{\cal D}_{520}+\frac{1}{32}{\cal D}_{521} + \frac{1}{64 F_\sigma^2 g^2} {\cal B}_{1}\notag\\
& + \frac{3}{64 F_\sigma^2 g^2} {\cal B}_{2} + \frac{7}{64 F_\sigma^2 g^2} {\cal B}_{3}\\
K_{22}^W \leftrightarrow & \frac{1}{16}{\cal D}_{511}+\frac{1}{8}{\cal D}_{512}+\frac{1}{8}{\cal D}_{520}+\frac{1}{16}{\cal D}_{521}- \frac{1}{8 F_\sigma^2 g^2} {\cal B}_{2}- \frac{1}{8 F_\sigma^2 g^2} {\cal B}_{3} \\
c_{1}^W \leftrightarrow &  - \frac{1}{4} {\cal D}_{516}\\
c_{2}^W \leftrightarrow & \frac{1}{8}{\cal D}_{496}+\frac{1}{16}{\cal D}_{497} - \frac{1}{4}{\cal D}_{499} - \frac{1}{4}{\cal D}_{500} - \frac{1}{4}{\cal D}_{517} - \frac{1}{4}{\cal D}_{518} - \frac{1}{2}{\cal D}_{519} \\
c_{3}^W \leftrightarrow & \frac{1}{2}{\cal D}_{499}+\frac{1}{2}{\cal D}_{500}+\frac{1}{4}{\cal D}_{518}+\frac{1}{2}{\cal D}_{519} \\
c_{5}^W \leftrightarrow &  - \frac{1}{4} {\cal D}_{516}\\
c_{6}^W \leftrightarrow &  - \frac{1}{8}{\cal D}_{497}+\frac{1}{4}{\cal D}_{500} - \frac{1}{4}{\cal D}_{501}+\frac{1}{4}{\cal D}_{517}+\frac{1}{2}{\cal D}_{519} - \frac{1}{32}{\cal D}_{520} \\
c_{7}^W \leftrightarrow &  - \frac{1}{2}{\cal D}_{500}+\frac{1}{4}{\cal D}_{501}+\frac{1}{16}{\cal D}_{509}+\frac{1}{32}{\cal D}_{511}+\frac{1}{16}{\cal D}_{512}+\frac{1}{16}{\cal D}_{513} - \frac{1}{2}{\cal D}_{519}\notag\\
&+\frac{3}{32}{\cal D}_{520}+\frac{1}{32}{\cal D}_{521} \\
c_{8}^W \leftrightarrow & \frac{1}{4}{\cal D}_{500} - \frac{1}{32}{\cal D}_{509} - \frac{1}{64}{\cal D}_{511} - \frac{1}{32}{\cal D}_{512} - \frac{1}{32}{\cal D}_{513}+\frac{1}{4}{\cal D}_{519} - \frac{3}{64}{\cal D}_{520} - \frac{1}{64}{\cal D}_{521} \\
c_{9}^W \leftrightarrow &  - \frac{1}{16}{\cal D}_{502} - \frac{1}{8}{\cal D}_{509} - \frac{1}{16}{\cal D}_{511}+\frac{1}{8}{\cal D}_{512} - \frac{1}{8}{\cal D}_{513}+\frac{1}{16}{\cal D}_{515} - \frac{1}{4}{\cal D}_{520} - \frac{1}{8}{\cal D}_{521} \\
c_{10}^W \leftrightarrow &  - \frac{1}{16}{\cal D}_{505} - \frac{1}{8}{\cal D}_{510} - \frac{1}{32}{\cal D}_{511} - \frac{1}{4}{\cal D}_{512} - \frac{1}{8}{\cal D}_{514}+\frac{1}{8}{\cal D}_{515}+\frac{11}{64}{\cal D}_{520} - \frac{1}{16}{\cal D}_{521} \\
c_{11}^W \leftrightarrow & \frac{1}{16}{\cal D}_{511}+\frac{1}{8}{\cal D}_{512}+\frac{1}{8}{\cal D}_{513}+\frac{1}{8}{\cal D}_{520}+\frac{1}{16}{\cal D}_{521} \\
c_{12}^W \leftrightarrow &  - \frac{1}{8}{\cal D}_{514} - \frac{1}{16}{\cal D}_{515} \\
c_{13}^W \leftrightarrow &  - \frac{1}{8}{\cal D}_{511} - \frac{1}{4}{\cal D}_{512} - \frac{1}{4}{\cal D}_{520} - \frac{1}{8}{\cal D}_{521}
\end{align}

 \bibliography{refs.bib}
\bibliographystyle{JHEP.bst}

 \end{document}